\documentclass[numberedappendix,twocolumn,twocolappendix]{openjournal}
\usepackage{float}
\usepackage{xcolor}
\usepackage{hyperref}
\hypersetup{
    unicode, 
    colorlinks=true,
    linkcolor=linkcolor,
    citecolor=linkcolor,
    filecolor=linkcolor,
    urlcolor=linkcolor,
}
\definecolor{linkcolor}{rgb}{0.0,0.3,0.5}

\usepackage{graphicx} 
\usepackage{amsmath} 
\usepackage{multirow} 
\usepackage{natbib} 
\usepackage{orcidlink}
\usepackage{subfigure}
\usepackage{afterpage}

\begin{document}

\title{An Extended Parametric Model for Self-Interacting Dark Matter Halos}

\author{Siddhesh M. Raut\orcidlink{0000-0002-3105-084X}}
\email{smraut@usc.edu}
\affiliation{Carnegie Observatories, 813 Santa Barbara St., Pasadena, CA 91101, USA}
\affiliation{Department of Physics $\&$ Astronomy, University of Southern California, Los Angeles, CA 90007, USA}

\author{Ethan O.~Nadler\orcidlink{0000-0002-1182-3825}}
\email{enadler@ucsd.edu}
\affiliation{Department of Astronomy \& Astrophysics, University of California, San Diego, La Jolla, CA 92093, USA}

\author{Andrew Benson\orcidlink{0000-0001-5501-6008}}
\email{abenson@carnegiescience.edu}
\affiliation{Carnegie Observatories, 813 Santa Barbara St., Pasadena, CA 91101, USA}

\begin{abstract}
    We improve upon the parametric model for the evolution of the density profiles of self-interacting dark matter (SIDM) halos introduced in \cite{yang2024parametric}, by considering the effects of mass accretion on a SIDM halo's gravothermal evolution. The original parametric model accurately predicts parameters $V_{\max}$ and $R_{\max}$, but with a tendency to overpredict $V_{\max}$ at $z=0$ for a subset of field halos. This discrepancy results from the parametric model predicting a faster rate of gravothermal evolution for these field halos compared to that measured in cosmological zoom-in simulations. We propose that the effects of mass accretion on the evolution of SIDM halos are not fully captured by the original parametric model. Our extended parametric model assumes that smooth mass accretion delays core-collapse by driving the SIDM halo back toward a Navarro-Frenk-White (NFW) profile (as it would have in the case of cold dark matter). We find that this extended model is able to substantially reduce the error in predicted $V_{\max}$ for halos compared to the original model, providing a more accurate model of SIDM halo evolution. 
\end{abstract}

\section{Introduction}

SIDM is a compelling model of dark matter proposed to remedy the shortcomings of the cold dark matter (CDM) model on small scales. These shortcomings originally included the too-big-to-fail \citep{boylan2011too_tbtf}, missing satellites \citep{klypin1999missing}, cusp--core \citep{moore1994evidence} problems, and diversity problems \citep{Oman_2015_diversity}. The missing satellites problem is now considered to be resolved thanks to new discoveries of faint satellites and improved theoretical predictions of stellar/mass luminosity relations \citep{Bullock_2017_small_scale_changes}. Similarly, recent work simulating halos with baryonic physics has also thrown the core--cusp tension into question \citep{Collins_2022}. The diversity problem, however, remains an important challenge for CDM \citep{sales2022baryonicsolutionschallengescosmological}. In CDM-only simulations, halos of all mass scales are predicted to roughly follow an NFW density profile \citep{navarro1995simulations_nfw,Navarro_1997_nfw_univ}. The cusp--core and diversity problems of CDM highlight the fact that not all observed halos have a dense, cuspy NFW profile. Specifically, it points out that there exist halos, of which most are dwarf halos, whose inner density profiles are much more `cored' than the cuspy inner profiles that the NFW profile suggests \citep{moore1994evidence}. The diversity problem further points out that there exists a general variance in density profiles that is not accounted for by CDM-only simulations \citep{Oman_2015_diversity}. Observations of disk galaxies show profiles that span the range from isothermal flat density cores to inner densities that are cuspier than NFW profiles \citep{Santos_Santos_2020, Ren_2019}. This range of density profile slopes is more prominent in halos that host galaxies with stellar masses in the range $10^5\mathrm{M}_\odot$--$10^7 \mathrm{M}_\odot$ \citep{Ren_2019}. 

The ``too-big-to-fail'' problem brings to light the fact that observations of the largest satellite galaxies of the Milky Way galaxy have circular velocities that do not align with expectations from CDM N-body simulations. Specifically, CDM N-body simulations of Milky Way-like halos produce subhalo populations that include many ($\sim 10$) dense subhalos that are high in mass and have $V_{\max} \gtrsim 30 \,\text{km s}^{-1}$, where $V_{\max}$ is the maximum circular velocity of a halo. Meanwhile, observations imply that the brightest Milky Way dSPhs have $V_{\max} <  30 \,\text{km s}^{-1}$ \citep{walker2009universal_dsphs_vmax}. Even after assigning the largest of the subhalos in the CDM simulations to host the Magellanic clouds and Sagittarius galaxy, there are still many subhalos with $V_{\max} \gtrsim 30 \,\text{km s}^{-1}$, but there are no observed satellite galaxies with $V_{\max} \gtrsim 30 \,\text{km s}^{-1}$ \citep{boylan2011too_tbtf}. If these large CDM halos did exist, they would be too massive \emph{not to} attract baryonic matter and host luminous galaxies, hence ``too big to fail''. There has been work done in the field that shows that taking into account baryonic physics can alleviate the ``too big to fail'' problem, since adding baryonic physics into simulations results in feedback and stronger tidal forces from disk galaxies in the host halo resulting in large subhalos whose profiles are more cored and less dense or even subhalos that are dark \citep{Wetzel_2016, Garrison_Kimmel_2019}. 

SIDM may provide a compelling solution to the core-cusp, diversity, and too-big-to-fail tensions. Due to the thermalization of halos arising from self-interactions between dark matter particles in SIDM, halos spend the first portion of their lifetime redistributing energy into the cold central cusp, resulting in the formation of flat isothermal cores as the self-interactions relax the inner core of the halo \citep{spergel2000observational, Tulin_2018}. Eventually, through further self-interactions, energy is transferred from the relatively hot central core to the cold outer regions of the halo. In this negative heat capacity environment, the interactions cause a significant loss of energy in the core, leading to contraction and a higher temperature as gravity concentrates more particles in the center \citep{balberg2002self_gravo_catastrophe, Essig_2019, Nishikawa_2020, turner2021onset}. This results in a runaway process known as ``core collapse'' that can produce density profiles even cuspier than the NFW profile's inner slope of $\rho \propto r^{-1}$. This diversity of density profile shapes throughout the gravothermal evolution of a SIDM halo warrants further investigation as a possible explanation for the tensions mentioned above. 

SIDM may also provide explanations for observations of more novel dark matter probes such as gravitational lensing and stellar streams. Gravitational lensing allows us to probe perturbations to the main lensed image caused by mass perturbations in the lens and along the line of sight. When our strong lens is a host halo, this technique allows us to probe subhalos that are mass perturbers affecting the strong lensed image. This allows us to probe dark matter substructure without having to rely on directly observing baryons from subhalos.  SDSS J0946+1006 (J0946) is a gravitational lens galaxy with a strong perturber of mass $M_\mathrm{sub} = (3.51 \pm 0.15) \times 10^9M_\odot$ \citep{Vegetti_2010_gravlens}. Further analysis by \cite{Minor_2021} of this perturber in J0946's lensed image show that the subhalo responsible for this must have a steep average log-slope of the density profile from between 0.75~kpc and 1.2~kpc of $\gamma_\mathrm{2D} < -1$ and a V-band luminosity of $L_\mathrm{V} <1.2 \times 10^8L_\odot$. They also perform forward analysis using the Illustris TNG100-1 simulation to show that CDM struggles to explain the existence of a subhalo with the properties mentioned prior \citep{Minor_2021}. \cite{Despali_2025,Tajalli_2025, enzi2025overconcentrated} all also find it unlikely for CDM to explain the existence of J0946's strong perturber; on the other hand, SIDM core-collapse can potentially explain the anomaly \citep{nadler2023GROUP}. In addition to motivating SIDM as a viable dark matter model, gravitational lensing can even be used to differentiate between SIDM models as shown in \cite{kong2025stronglensingperturberssidm}.  

Initial work with stellar streams have also found gaps and spurs that can be explained by high density compact objects like small high concentrated subhalos. \cite{bonaca2019spur} and \cite{nibauer2025measurementdarkmattersubstructure} both find that the compact dense object responsible for the spur-and-gap features for the GD-1 stream has a mass in the range of $10^{5.5} - 10^8 M_\odot$ with inferred densities higher than CDM predictions. \cite{zhang2024gd1stellarstreamperturber} shows that while forward modeling cosmological zoom in simulations of CDM dark matter fail to produce a dense and compact enough subhalo to explain the spurs and gaps in GD-1, isolated SIDM n-body simulations manage to produce core collapsed halos with inner densities high enough and compact to be GD-1's perturber. Dense, core‑collapsed SIDM halos can explain the compact dark perturbers seen in a stellar stream, a strong lens, and a Milky Way satellite, tying three small‑scale dark‑matter probes to SIDM.\citep{yu2025birdsstonecorecollapsedsidm}.

Early research on SIDM focused on constant scattering cross sections, $\sigma/m$, constrained to match observations. However, observations of galaxy clusters and dwarf satellite galaxies yielded incompatible cross-section constraints. Galaxy cluster observations yielded constraints of $\sigma / m \lesssim  0.1 \,\mathrm{cm^2/g}$ at particle relative velocity scales of $V \approx 1000 \,\mathrm{km/s}$ \citep{kaplinghat2016dark}, while observations of dwarf satellites suggested that $\sigma / m \geq 1 \,\mathrm{cm^2/g}$ at $V \lesssim 100 \,\mathrm{km/s}$ \citep{zavala2013constraining_dwarf}.

This tension between different velocity scales can be alleviated via a velocity-dependent cross section \citep{silverman2023motivations, kaplinghat2016dark, kamada2020escalating, engelhardt2026marvelouslydarkgravothermalevolution}. This allows dark matter particles to have a lower cross section at the high relative velocities typical of clusters. In contrast, at lower mass scales, where relative velocities are smaller, dark matter can have larger cross sections. This allows the effects of self-interactions to be heightened at specific scales and suppressed at others. Several velocity-dependent cross sections have been proposed, ranging from power law dependencies, $\sigma \propto v^{-n}$, to Yukawa-type scattering relations \citep{yang2022gravothermal, Sagunski_2021}. In this work, we will consider SIDM cross-sections with a velocity-dependent Rutherford scattering model \citep{nadler2023GROUP,yang2022gravothermal, ibe2010distinguishing}:

\begin{equation}
    \frac{\mathrm{d}\sigma}{\mathrm{d}\cos\theta} = \frac{\sigma_0}{2\left[ 1 + \frac{\upsilon^2}{w^2} \sin^2(\theta / 2)  \right]^2},
    \label{eq:sidmcross}
\end{equation}
where $\sigma_0$ controls the cross section in the low velocity limit, $w$ is a parameter that controls the velocity scale at which the cross section begins to decrease as $v^{-4}$, and $\theta$ is the scattering angle.

When modeling velocity-dependent SIDM with N-body simulations, running multiple cosmological zoom-in simulations to compare to observations for every set of model parameters in this cross-section parametrization is impractically expensive. However, there are analytic models that offer a fast alternative to predict how isolated halos evolve in SIDM and CDM models. The parametric model, introduced by \cite{yang2024parametric, yang2024testing}, allows a population of halos with density profiles appropriate to any given SIDM cross section to be generated using the formation histories of halos extracted from a single CDM cosmological zoom-in simulation, avoiding the need to run an entirely new N-body simulation for each SIDM cross section to be considered.

In this work, we improve upon the accuracy of the parametric model of \cite{yang2024parametric} by taking into account the effects of mass accretion on the gravothermal evolution of SIDM halos. This paper is organized as follows. Section \ref{section:methods} presents the details of the N-body simulations used in this work and explains how the original and extended parametric models are applied to this data. Sections \ref{sec:results_group} and \ref{sec:results_MW} analyze how our extended analytic model performs relative to the original formulation, highlighting improvements in the accuracy of SIDM halo density profile predictions. Finally, in Section~\ref{sec:conclusions} we present our conclusions.


\section{Methods}\label{section:methods}

\subsection{Simulations} \label{section:simulation}

This work utilizes two pairs of zoom-in simulations corresponding to Group \citep{nadler2023GROUP} and Milky Way \citep{yang2023MW_diversify} halo environments. These halos are part of the SIDM Concerto suite \citep{Nadler_2025_concerto} and will be referred to as the ``Group'' and ``Milky Way'' simulations, respectively, hereafter. Each pair consists of two N-body cosmological zoom-in simulations run with two different dark matter models: an SIDM and a CDM model. For a given pair, the two simulations are initialized with the same initial conditions, allowing any differences between the simulations to be attributed to the differences in the dark matter models used to run the simulations. In each pair, a population of CDM halos is extracted from the CDM simulation, from which the parametric models (to be described in Section \ref{section:param_model}) can be used to predict a corresponding population of SIDM halos. The analogous population of SIDM halos, extracted from the SIDM simulation in each pair, is then used to check how closely the parametric models match the simulation results. All simulations used in this work were carried out using \textsc{Gadget-2} \citep{Springel_2005_Gadget-2}. The SIDM simulations employed code developed by \cite{yang2023MW_diversify} and \cite{yang2022gravothermal}, which is implemented within \textsc{Gadget-2}. 

Equation~\ref{eq:sidmcross} gives the differential cross section used in the SIDM simulations. Both SIDM simulations in this work, by design, have $w$ chosen such that a fraction of the subhalos and field halos in each simulation will core collapse by $z=0$. 


\begin{table*}
    \centering
    \caption{Properties of the simulations used in this work. Columns give the name of the simulation group, the particle mass, the gravitational softening length, and SIDM cross-section parameters as used in the two pairs of simulations.}
    \label{table:basicparam}
    \begin{tabular}{llrrr}
     \hline
     & & & \multicolumn{2}{c}{\textbf{SIDM cross section}} \\
     \textbf{Simulation} & \textbf{Particle mass} &  \textbf{Softening length [{\boldmath $h^{-1}$}~pc]} & {\boldmath $\sigma_0/m$} \textbf{[cm$^2$/g]} & {\boldmath $w$} \textbf{[km/s]} \\
     \hline
     Group     & $3 \times 10^5 \ h^{-1} \ \mathrm{M}_\odot$   & $170$  & 147.1   & 120.00\\
     Milky Way & $4 \times 10^4 \ h^{-1} \ \mathrm{M}_\odot$   & $80$  & 147.1   & 24.33\\
     \hline
    \end{tabular}
\end{table*}

\subsubsection{Group simulations}
The Group pair of simulations are SIDM and CDM cosmological zoom-in simulations focused on a host halo of $M_\mathrm{vir}(z=0) = 1.1 \times 10^{13} \ \mathrm{M}_\odot$ whose initial conditions are taken from the ``Group'' suite of Symphony simulations \citep{Nadler_202_symphony, nadler2023GROUP}. Table \ref{table:basicparam} shows the particle mass and force softening length used for these simulations, along with the SIDM cross-section parameters as defined in Equation~(\ref{eq:sidmcross}). 

The $\sigma_0$ and $w$ parameters used in the Group simulation are chosen such that halos in the simulation, which span a range of $30~\mathrm{km}~\mathrm{s}^{-1} \lesssim V_\mathrm{max} \lesssim  300~\mathrm{km}~\mathrm{s}^{-1}$, have corresponding effective cross-sections in the range of $ 2~\mathrm{cm}^2~\mathrm{g}^{-1} \lesssim \sigma_\mathrm{eff}/m \lesssim 100~\mathrm{cm}^2~\mathrm{g}^{-1}$. As defined in \cite{yang2022gravothermal}, the effective cross-section is a velocity-weighted average over the velocity-dependent differential cross-section defined as:
\begin{equation}
    \sigma_{\text{eff}} = 
    \frac{
    2 \int \mathrm{d}v \mathrm{d}\cos\theta \, \frac{\mathrm{d}\sigma}{\mathrm{d}\cos\theta} \sin^2\theta \, v^5 f_{\mathrm{MB}}(v, \nu_\mathrm{eff})
    }{
    \int \mathrm{d}v \cos\theta \sin^2\theta \, v^5 f_{\mathrm{MB}}(v, \nu_\mathrm{eff})
    }, \label{eq:effective-cross}
\end{equation}
where $f_\mathrm{MB}(v, \nu_\mathrm{eff})$ is the distribution of velocity for particles in our given halo, which we have assumed to follow a Maxwellian distribution:

\begin{equation}
    f_{\mathrm{MB}}(v, \nu_\mathrm{eff}) \propto v^2 \exp \left[ - \frac{v^2}{4\nu^2_\mathrm{eff}}\right],
\end{equation}
for which $\nu_\mathrm{eff} = 0.64V_{\max}$. In the velocity-dependent regime, lower-mass (sub)halos have higher $\sigma_\mathrm{eff}/m$, resulting in higher interaction rates and core collapse at earlier times, while higher-mass halos evolve much more slowly due to their lower $\sigma_\mathrm{eff}/m$. 

\subsubsection{Milky Way simulations}

The second pair of simulations used in this work are the Milky Way simulations \citep{yang2023MW_diversify}, originally drawn from \cite{Mao_2015} and later incorporated into the Symphony Milky Way suite \citep{Nadler_202_symphony}. This pair of simulations are also cosmological zoom-ins, but are zoomed in on a Milky Way-like system chosen to have a merger history similar to that of the actual Milky Way \citep{yang2023MW_diversify}. This system has a host halo with comparable mass to that of the Milky Way, $M_\mathrm{vir}(z=0) = 1.6 \times 10^{12} \ \mathrm{M}_\odot$, and also has a satellite halo analogous to the LMC, which accretes into the Milky Way halo approximately 1~Gyr before the present time. The scattering model for the SIDM simulation of this pair follows Equation~\eqref{eq:sidmcross} with $ \sigma_0 = 147.1 \ \text{cm}^2 \ \text{g}^{-1} $ and $ w = 24.33 \ \text{km} \ \text{s}^{-1} $ as shown in Table \ref{table:basicparam}. Similar to the Group simulations, the $w$ parameter is chosen to allow smaller halos---in particular, subhalos of the Milky Way-sized host halo---to core collapse by $z=0$.

\subsubsection{Halo Samples}


For each pair of simulations, we must select a population of field halos on which to test our models. In this work, we use \textsc{Rockstar} \citep{rockstar} and \textsc{Consistent-Trees} \citep{consistent_trees} to identify halos and subhalos and to measure relevant halo properties for each snapshot for each of the four simulations.

To select a population of field halos with which we can reliably test the parametric models used in this work, we apply a series of cuts on the population of halos in the CDM simulation at $z=0$ to create our sample of CDM field halos.  Using the data from \textsc{Rockstar} and \textsc{Consistent-Trees}, we first apply a cut that retains only those halos that have been exclusively field halos since the time at which they first assembled half of their $z=0$ mass. This ensures that backsplash subhalos do not contaminate our sample of halos. Next, we apply a mass cut, identical to that used by \cite{yang2024testing} (and corresponding to 4,000 particles), to remove halos that may be affected by the simulation resolution. Finally, we exclude halos that may be strongly affected by tidal forces from the largest halo in the simulation and exclude halos outside of the zoom-in region that may contain lower resolution particles by removing halos outside of a range of distance from the central host halo. We adopt values for these mass and distance selections that are similar to those used by \cite{yang2024testing}, allowing for comparisons between this work and theirs. For the Milky Way CDM simulation, once the cut for field halos has been applied, halos are retained if their virial mass $M_\mathrm{vir} > 1.43 \times 10^8 \mathrm{M}_\odot$, and they reside within 0.3--3~Mpc of the central Milky Way halo. For the Group CDM simulation, halos are retained if $M_\mathrm{vir} > 6 \times 10^9 \mathrm{M}_\odot$ and they are within 0.6--8~Mpc of the central group halo.

Once this is done, for each CDM halo in this sample, we find the matching SIDM halo in the corresponding SIDM simulation to create the sample of SIDM halos. Since we are focused on field halos surrounding the largest host halo in each simulation, we are able to match halos between the SIDM and CDM realizations in each pair of simulations by using each field halo's distance to the center of the main host halo. After calculating the CDM halo's distance from the main host halo in the simulation as a function of scale factor, we apply the same cuts as mentioned above to SIDM halos, except for the cut on the virial mass of the halo; this cut is excluded because it might create problems for finding matches for the lowest mass CDM halos in our sample if their SIDM counterparts experienced more mass loss and thus fall below the virial mass cut. Next, the relative distance of all SIDM halos that have survived our cuts are calculated. Finally, we find the matched SIDM halo by identifying the SIDM halo in our sample with the lowest mean absolute error in relative distance to the CDM halo.


\subsection{Parametric Models for SIDM Density Profiles} \label{section:param_model}

\cite{yang2024parametric} developed a parametric model which, given the formation history of a CDM halo, aims to predict the density profile of the corresponding SIDM halo. To generate entire populations of SIDM halos for a given SIDM particle model, the integral-based\footnote{\cite{yang2024parametric} introduce a ``basic'' and ``integral'' method for the parametric model. In this paper, we will use the integral method of the original parametric model. The new, ``extended'' model that we will introduce is also formulated on this integral-based parametric model.} parametric model from \cite{yang2024parametric} requires three inputs:
\begin{enumerate}
    \item a CDM cosmological N-body simulation of a population of halos;
    \item measurements of the properties of those CDM halos, including $V_{\max}$ and $R_{\max}$ for each halo at each snapshot;
    \item the cross-section of the SIDM model.
\end{enumerate}
In this work, halos are identified, and their properties are measured using the \textsc{Rockstar} halo finder and the \textsc{Consistent-Trees} merger tree algorithm. Of course, other methods could be used\footnote{In fact, any method that produces halo formation histories with the relevant properties---for example, merger trees constructed using extended Press-Schechter methods \citep{2000MNRAS.319..168C}---could be used to produce inputs to the parametric model.} providing they supply the required properties of each halo across its entire formation history. In the case of velocity-dependent differential cross-sections, $\mathrm{d}\sigma/\mathrm{d}\cos(\theta)$, there is no one single cross-section for the entire SIDM model. Instead, an effective cross-section is calculated for use with the parametric model using Equation \ref{eq:effective-cross}.
 
In the remainder of this section, we first summarize the original parametric model presented in \cite{yang2024parametric} and then describe the extensions we make to it to achieve improved agreement with simulation results. 

\subsubsection{Original parametric model} \label{section:original}
The original parametric model \citep{yang2024parametric} takes a few key properties of halos from a cosmological zoom-in N-body CDM simulation, measured across a halo's evolution, and uses two key integrals to predict how the density profile of an analogous SIDM halo in the same environment would evolve. Specifically, $V_{\max}$, the maximum circular velocity in the halo's rotation curve, and $R_{\max}$, the radius at which that maximum velocity occurs, are chosen as the two parameters to track the evolution of both the measured CDM halo and the predicted SIDM halo because they are profile-independent parameters that fully characterize the structure of a CDM halo (assuming it follows the NFW form).

The original parametric model maps the evolution of $V_{\max}$ and $R_{\max}$ from a CDM halo onto a predicted SIDM halo using the following two integrals:
\begin{eqnarray}
    V^\mathrm{(SIDM)}_\mathrm{max}(t) &=& V^\mathrm{(CDM)}_\mathrm{max}(t_\mathrm{f}) + \int_{t_\mathrm{f}}^{t} \mathrm{d}t^\prime \frac{\mathrm{d}V^\mathrm{(CDM)}_\mathrm{max}(t^\prime)}{\mathrm{d}t^\prime}  \nonumber \\ 
    & & +  \int_{t_\mathrm{f}}^{t} \frac{\mathrm{d}t^\prime}{t_\mathrm{c}(t^\prime)} \frac{\mathrm{d}V^\mathrm{(model)}_\mathrm{max}(\tau')}{\mathrm{d}\tau'},
    \label{eq:vmaxintegral}
\end{eqnarray}
and
\begin{eqnarray}
    R^\mathrm{(SIDM)}_\mathrm{max}(t) &=& R^\mathrm{(CDM)}_\mathrm{max}(t_\mathrm{f}) + \int_{t_\mathrm{f}}^{t} \mathrm{d}t^\prime \frac{\mathrm{d}R^\mathrm{(CDM)}_\mathrm{max}(t^\prime)}{\mathrm{d}t^\prime} \nonumber \\
    & & + \int_{t_\mathrm{f}}^{t} \frac{\mathrm{d}t^\prime}{t_\mathrm{c}(t^\prime)} \frac{\mathrm{d}R^\mathrm{(model)}_\mathrm{max}(\tau')}{\mathrm{d}\tau'}.
    \label{eq:rmaxintegral}
\end{eqnarray}
Here, $\tau = (t-t_\mathrm{f})/t_\mathrm{c}$ is a measure of how far our SIDM halo has evolved toward core collapse, $t_\mathrm{c}$ is the timescale to core collapse for a given SIDM halo as defined by \citep{Essig_2019,yang2024parametric}:
\begin{equation}
    t_\mathrm{c} = \frac{150}{C} \frac{1}{(\sigma_{\mathrm{eff}}/m)\rho_{\mathrm{eff}}r_{\mathrm{eff}}} \frac{1}{\sqrt{4\pi \mathrm{G} \rho_{\mathrm{eff}}}},
    \label{eq:tc}
\end{equation}
where 
\begin{align}
    \rho_\mathrm{eff} &= G^{-1} \left(\frac{V_\mathrm{max}}{1.648 r_\mathrm{eff}}\right)^2,\label{eq:rho_eff} \\ 
    r_\mathrm{eff} &= \frac{R_\mathrm{max}}{2.1626}, \label{eq:r_eff}
\end{align}
for any given halo, and $t_\mathrm{f}$ is the time at which the halo is formed. Details of how $t_\mathrm{f}$ is calculated in the parametric model can be found in Appendix \ref{sec:formation_time_app}. 

The dimensionless time parameter, $\tau$, can exceed the value of 1, but we limit the range to $0 \leq \tau \leq 1$ where $\tau = 0.15$ corresponds to the time when the maximum core size is achieved and $\tau = 0.75$ corresponds to the stage when the SIDM central density first exceeds CDM \citep{Roberts_2025,Nadler_2025_cozmic}.  It is important to mention that in Equations \ref{eq:vmaxintegral} and \ref{eq:rmaxintegral}, $V^\mathrm{(CDM)}_\mathrm{max}$ and $R^\mathrm{(CDM)}_\mathrm{max}$ are actual parameters of the CDM halo in the N-body simulation that are calculated using \textsc{Rockstar}; $V^\mathrm{(SIDM)}_\mathrm{max}$ and $R^\mathrm{(SIDM)}_\mathrm{max}$, on the other hand, are values being generated by the parametric model to emulate a hypothetical SIDM halo. 

Both integrals, Equations~\ref{eq:vmaxintegral} and \ref{eq:rmaxintegral}, work in the same manner but are applied to two different parameters, $V_{\max}$ and $R_{\max}$, respectively. The first two terms in each integral simply recreate $V^\mathrm{(CDM)}_\mathrm{max}(t)$ or $R^\mathrm{(CDM)}_\mathrm{max}(t)$. The third term in each integral takes into account the SIDM physics. In order to predict how much $V_{\max}$ and $R_{\max}$ change due to SIDM physics, the third integral integrates over a model function with respect to $\tau$.  \cite{yang2024parametric} show that $V^\mathrm{(SIDM)}_\mathrm{max}(t)/V^\mathrm{(CDM)}_\mathrm{max}(t)$ and  $R^\mathrm{(SIDM)}_\mathrm{max}(t)/R^\mathrm{(CDM)}_\mathrm{max}(t)$, as determined from a set of isolated N-body SIDM halo simulations, which follow the full evolution of a SIDM halo from an initial NFW profile through core formation and all the way to core collapse, can be fit by functions of $\tau$:
\begin{eqnarray}
\frac{V^\mathrm{(SIDM)}_\mathrm{max}(t)}{V^\mathrm{(CDM)}_\mathrm{max,0}} &=&
1 + 0.1777\tau - 4.399\tau^3 + 16.66\tau^4 - 18.87\tau^5 \nonumber \\
 & & + 9.077\tau^7 - 2.436\tau^9,
\label{eq:param_Vmax_cal}
\end{eqnarray}

\begin{equation}
\frac{R^\mathrm{(SIDM)}_\mathrm{max}(t)}{R^\mathrm{(CDM)}_\mathrm{max, 0}} =
1 + 0.007623\tau - 0.7200\tau^2 + 0.3376\tau^3 - 0.1375\tau^4.
\label{eq:param_Rmax_cal}
\end{equation}

They further show that, in these same simulations, the structural parameters of the SIDM halo (density normalization, scale, and core radii) can also be expressed in terms of the density normalization and scale radius of their CDM counterpart halo, and $\tau$.

The parametric model starts at $t_\mathrm{f}$ and evaluates Equations~\ref{eq:vmaxintegral} and \ref{eq:rmaxintegral} at every timestep, $t$. For every iteration, the parametric model first calculates $t_\mathrm{c}(t)$ by substituting $V^\mathrm{(CDM)}_\mathrm{max}(t)$ and $R^\mathrm{(CDM)}_\mathrm{max}(t)$ into Equations~\ref{eq:tc}--\ref{eq:r_eff} as previously described. With, $t_\mathrm{c}(t)$ and $t_\mathrm{f}$, we calculate $\tau(t)$ at the timestep, $t$. With $\tau(t)$ calculated for the current timestep, the parametric model calculates $V^\mathrm{(SIDM)}_\mathrm{max}$ and $R^\mathrm{(SIDM)}_\mathrm{max}$ with Equations~\ref{eq:vmaxintegral} and \ref{eq:rmaxintegral}.

At this point, the parametric model has predicted the profile-independent parameters, $V^\mathrm{(SIDM)}_\mathrm{max}(t)$ and $R^\mathrm{(SIDM)}_\mathrm{max}(t)$, and all that is left to do is calculate the SIDM profile-dependent parameters: $\rho_\mathrm{s}$, $r_\mathrm{s}$, and $r_\mathrm{c}$ that describe the SIDM density profile using the functional form:

\begin{equation}
    \rho_{\text{SIDM}}(r) = \frac{\rho_\mathrm{s}}{[(r^\beta + r_\mathrm{c}^\beta)^{1/\beta}/r_\mathrm{s}](r/r_\mathrm{s} + 1)^2},
    \label{eq:sidm_profile}
\end{equation}
where $\beta = 4$ \citep{yang2024parametric}.

This is done by calibrating equations similar to Equations~\ref{eq:param_Vmax_cal} and \ref{eq:param_Rmax_cal} with $\tau$. The details of how this is done are described in Appendix \ref{sec:SIDM_param_app}, since this part of the parametric model remains unchanged from the original model described in this section through the extended parametric model presented in the next section.
\vspace{\baselineskip}
\subsubsection{Extended Parametric Model} \label{section:Mass accretion model}

The original parametric model works well for many halos, but we find that there is a fraction of field halos for which the original model overpredicts the rate at which the halo core collapses. For these halos, the original model predicts a faster growth of $V_{\max}$ than the actual SIDM N-body simulation results. In this work, we hypothesize that the original parametric model does not fully capture the effects of mass accretion throughout the formation history of halos. Mass accretion onto a halo adds energy and particles to the outskirts of the halo. This influx of heat into the halo introduces extra energy that can be redistributed to the core, keeping the center of the halo in a core formation stage, thus delaying core collapse. We aim to capture this delaying effect of mass accretion on the evolution of a SIDM halo by introducing a simple way to incorporate the effects of mass accretion into $\tau$. 

To do this, we include mass accretion as part of the energy equation that governs the core collapse time scale. We begin by noting that $t_\mathrm{c}$ can be thought of as the timescale for self-interactions to transport the entire binding energy of the halo core to the outer regions. Furthermore, given that $\tau$ ranges from $0 \leq \tau \leq 1$ and represents the stage of core evolution in a SIDM halo, we can write:

\begin{align}
    t_\mathrm{c}(t) &\approx \frac{E_{\mathrm{SIDM}}(t)}{\dot{E}(t)}, \label{eq:tc_1stdef}\\
    \dot{\tau}(t) &\approx \frac{1}{t_\mathrm{c}},
    \label{eq:Edot_tau}
\end{align}
where $ \dot{E}(t)$ is the \emph{difference} in rate of change of energy (per unit volume) between SIDM and CDM halos, $ \dot{E}(t) = \dot{E}_{\mathrm{SIDM}}(t) - \dot{E}_\mathrm{CDM}(t) $. This tells us that the rate of increase of $\tau$ is proportional to the rate of energy transfer out of the SIDM halo compared to its CDM counterpart. Relating $\tau$ to the difference also allows us to later construct a way to connect the evolution between SIDM and CDM halos. Intuitively, it also makes sense since the more rapidly the energy is transferred out of the core of a SIDM halo due to self-interactions in comparison to its CDM counterpart, the faster we expect the core to evolve. Thus, the faster $\tau$ should increase. 

We can break $ \dot{E}(t)$ into a sum of cooling (due to SIDM interactions) and heating (due to mergers) terms:
\begin{equation}
    \dot{E}(t) = \mathcal{L}_\mathrm{SIDM} + \mathcal{H}_\mathrm{SIDM} - \mathcal{H}_\mathrm{CDM},
\end{equation}
where $\mathcal{L}_\mathrm{SIDM}$ is the heat transfer per unit volume from the SIDM halo due to self-interactions. This term accounts for the thermal effect of SIDM self-scattering, which allows SIDM halos to redistribute their energy to and from the core. $\mathcal{H}_\mathrm{SIDM}$ and $\mathcal{H}_\mathrm{CDM}$ are the heating rates due to mass being added from the environment for SIDM and CDM halos, respectively. They can be thought of as the rate at which the gravitational potential energy changes due to the addition of new mass:
\begin{align}
    \mathcal{H}_\mathrm{SIDM}  &=  \phi_\mathrm{SIDM} \dot{\rho}_\mathrm{SIDM}, \\
    \mathcal{H}_\mathrm{CDM}  &=  \phi_\mathrm{CDM} \dot{\rho}_\mathrm{CDM},
\end{align}
where $\phi$ is the gravitational potential of the halo.

This allows $\dot{E}(t)$ to be expressed as
\begin{align}
    \dot{E}(t) &= \mathcal{L}_\mathrm{SIDM} + \phi_\mathrm{SIDM} \dot{\rho}_\mathrm{SIDM} - \phi_\mathrm{CDM} \dot{\rho}_\mathrm{CDM} .
\end{align}

To first order, we assume that the rate of change in density is proportional to the rate of change of the mass that is being brought into the halo:
\begin{equation}
    \frac{\dot{\rho}}{\rho} = \frac{\dot{M}}{M}.
\end{equation}

To further simplify, we assume that $\dot{\rho}$ at any given moment should be the same for a given pair of CDM and SIDM halos. Since they are matched pairs, they exist in very similar environments and thus have the same mass accretion at some large radius; therefore, $\dot{\rho}_\mathrm{SIDM} = \dot{\rho}_\mathrm{CDM}$. However, we expect the gravitational potentials of our two paired CDM and SIDM halos to differ based on how far the SIDM halo has evolved; thus, $\phi_\mathrm{SIDM} \neq \phi_\mathrm{CDM}$. $\mathcal{L}_\mathrm{SIDM}$ can also be simplified to $\mathcal{L}_\mathrm{SIDM} = \phi_\mathrm{SIDM} \rho_\mathrm{SIDM}/t_\mathrm{c}$. Putting this together, we find:

\begin{align}
    \dot{E}(t) &= \frac{\phi_\mathrm{SIDM} \rho_\mathrm{SIDM}}{t_\mathrm{c}} + \phi_\mathrm{SIDM} \dot{\rho}_\mathrm{SIDM} - \phi_\mathrm{CDM} \dot{\rho}_\mathrm{CDM}, \\ 
    \dot{E}(t) &= \phi_\mathrm{SIDM} \rho_\mathrm{SIDM} \left( \frac{1}{t_\mathrm{c}} + \frac{\dot{\rho}_\mathrm{SIDM}}{\rho_\mathrm{SIDM}} - \frac{\phi_\mathrm{CDM} \dot{\rho}_\mathrm{SIDM}}{ \phi_\mathrm{SIDM} \rho_\mathrm{SIDM}} \right), \\
    \dot{E} &= \phi_\mathrm{SIDM} \rho_\mathrm{SIDM} \left( \frac{1}{t_\mathrm{c}} + \frac{\dot{M}}{M} -  \frac{\dot{M}}{M} \frac{\phi_\mathrm{CDM}}{\phi_\mathrm{SIDM}} \right).
\end{align}


We posit that $\phi_\mathrm{CDM}$ is related to $\phi_\mathrm{SIDM}$ through a function of $\tau$ alone:

\begin{align}
    \phi_\mathrm{CDM} = f(\tau)\phi_\mathrm{SIDM},
    \label{eq:potential_relation}
\end{align}
such that
\begin{align}
    \dot{E} &= \phi_\mathrm{SIDM} \rho_\mathrm{SIDM} \left( \frac{1}{t_\mathrm{c}} + \frac{\dot{M}}{M} -  \frac{\dot{M}}{M}f(\tau) \right).
\end{align}

\begin{figure*}[t]
\centering
\begin{tabular}{cc}
    \includegraphics[width=0.45\linewidth]{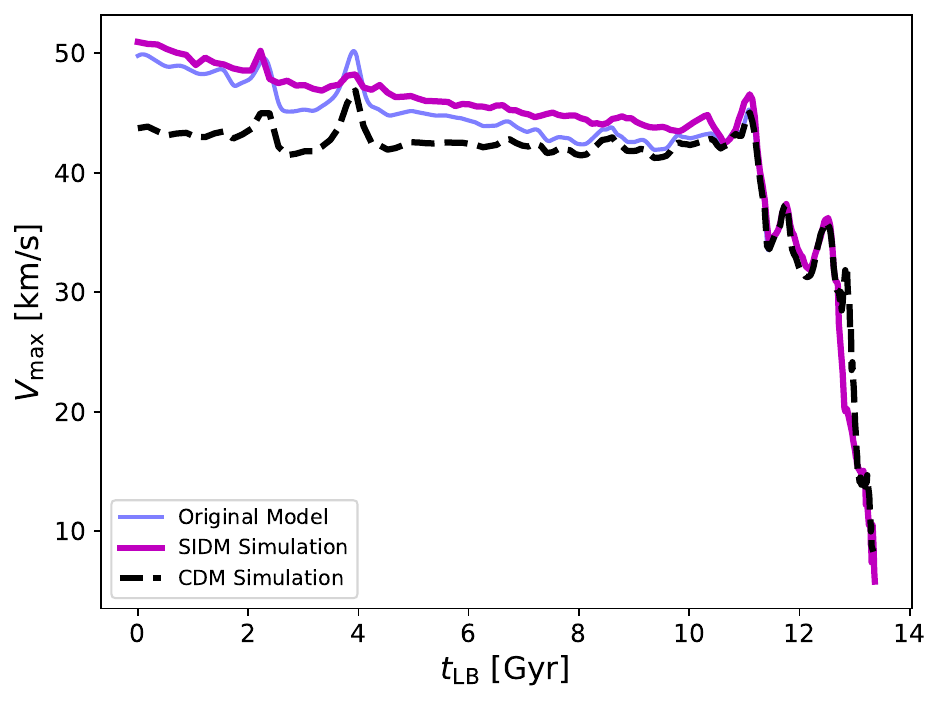}  &  \includegraphics[width=0.45\linewidth]{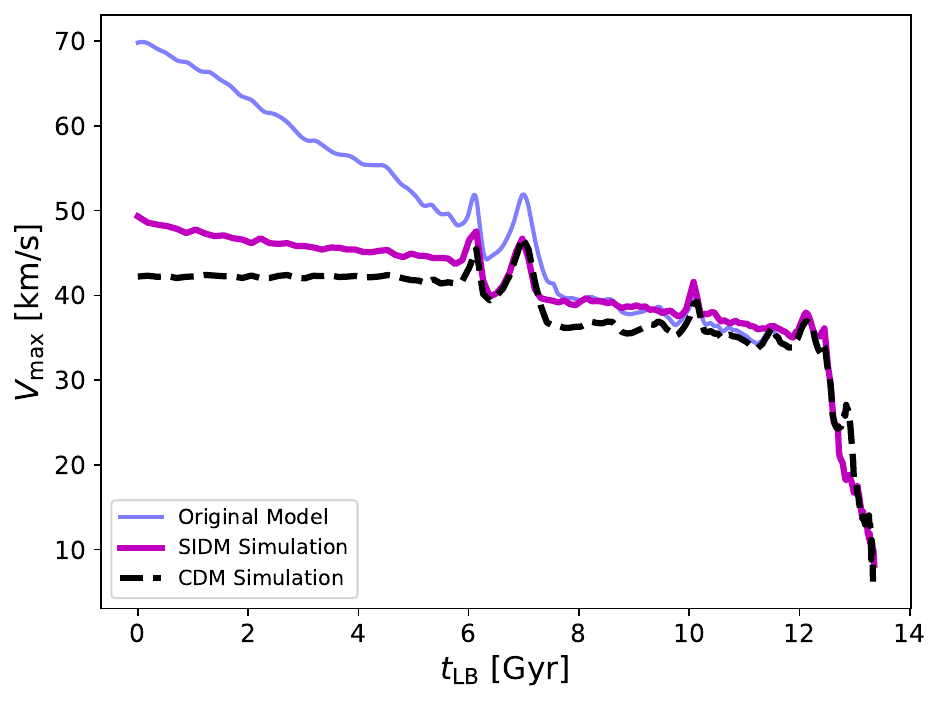} 
\end{tabular}
\caption{Examples of the evolution of $V_\mathrm{max}$ in SIDM halos as a function of lookback time, shown by purple lines. Black dashed lines show the evolution of $V_\mathrm{max}$ in the matched CDM halo, while the blue line indicates the predictions of the original parametric model, computed using the assembly history of the matched CDM halo. Left and right panels show cases where the original parameter model provides an accurate match to the SIDM simulation and significantly overpredicts $V_\mathrm{max}$, respectively.} 
\label{fig:Vmax_examples}
\end{figure*}

$f(\tau)$ can be Taylor expanded to first order $f(\tau) = f^{(0)}+f^{(1)}\tau+\ldots$ resulting in:

\begin{align}
    \dot{E} &= \phi_\mathrm{SIDM} \rho_\mathrm{SIDM} \left( \frac{1}{t_\mathrm{c}} + \frac{\dot{M}}{M} -  \frac{\dot{M}}{M} \left[f^{(0)}+f^{(1)}\tau \right] \right).
    \label{eq:simplified_Edot}
\end{align}

Looking back to Equation \ref{eq:potential_relation}, at $\tau = 0$ there has been no evolution by definition so $f(\tau = 0) = 1$. As a result $f^{(0)} = 1$. This simplifies Equation \ref{eq:simplified_Edot} to:

\begin{equation}
    \dot{E} = \phi_\mathrm{SIDM} \rho_\mathrm{SIDM} \left( \frac{1}{t_\mathrm{c}} - \frac{\dot{M}}{M} f^{(1)}\tau \right).
    \label{eq:final_Edot}
\end{equation}

Simply relabeling some of the variables in Equation \ref{eq:final_Edot} will give us the form of the equation we used to determine $\tau$ in our version of the parametric model: $\dot{M}/M = \Gamma$ and $f^{(1)} = \alpha$:
\begin{equation}
    \dot{E} = \phi_\mathrm{SIDM} \rho_\mathrm{SIDM} \left( \frac{1}{t_\mathrm{c}} - \Gamma \alpha \tau \right).
    \label{eq:tau_model}
\end{equation}

Recognizing that $\phi_\mathrm{SIDM} \rho_\mathrm{SIDM}$ is simply $E_{\mathrm{SIDM}}(t)$, the total energy per unit volume of our SIDM halo, we can then compare Equation \ref{eq:tau_model} with Equations~\ref{eq:tc_1stdef} and \ref{eq:Edot_tau} making it apparent that 

\begin{equation}
   \dot{\tau} = \frac{1}{t_\mathrm{c}} - \Gamma \alpha \tau.
    \label{eq:tau_dot}
\end{equation}

We use Equation \ref{eq:tau_dot} to calculate the change in $\tau$ at each timestep for the SIDM halo in the parametric model and update $\tau$. This is different than the original parametric model where $\tau = (t-t_\mathrm{f})/t_\mathrm{c}$. In fact, this is the only difference between the original parametric model and our extended parametric model. Other than how $\tau$ is calculated at each timestep, the extended parametric model follows the same steps and procedures outlined in Section \ref{section:original}. 

When running the extended parametric model, Equation \ref{eq:tau_dot} is calculated by using halo parameters provided by \textsc{Rockstar} and \textsc{Consistent-Trees} to calculate $t_\mathrm{c}$ and $\Gamma$ for each time step. Specifically, we take the $M_\mathrm{vir}$ data of each halo across snapshots provided by \textsc{Rockstar} and \textsc{Consistent-Trees} and create a smooth piece-wise interpolation using \texttt{interp1d} from the SciPy library. With this interpolated function of $M_\mathrm{vir}$, we then calculate a finite difference calculation as such: 

\begin{equation}
    \Gamma(t_i) = \frac{M(t_i + \Delta t/2) - M(t_i - \Delta t/2)} { \Delta t \cdot M(t_i - \Delta t/2)}
\end{equation}
to obtain the $\Gamma$ at timestep, $t_i$. At each timestep we use Equation~\ref{eq:tau_dot} to calculate $\mathrm{d}\tau$ as:

\begin{equation}
    \mathrm{d}\tau(t_i) = \left( \frac{1}{t_c(t_i)} - \alpha \Gamma(t_i) \tau(t_{i-1}) \right) \mathrm{d}t,
    \label{eq:dtau_discrete}
\end{equation}
where $\alpha$ is a constant of proportionality we keep fixed for the extended model, and $\Delta t$ is the time between each timestep.
The difference in calculating $\tau$ between our extended model and the original is significant. In the original model, at each timestep $\tau$ is calculated separately with no dependence on the previous timestep. In our model, not only do we look at mass from the halo before the current timestep to calculate $\Gamma(t_i)$, we also use $\tau(t_{i-1})$ from the previous timestep in order to calculate the amount of $\mathrm{d}\tau(t_i)$ we must add. Our change to how $\tau$ is calculated in the extended parametric model results in two major changes: 1) $\tau$ now depends on the mass accretion rate $\Gamma$ and 2) we approach calculating $\tau$ by adding increments of $\mathrm{d}\tau$, which we will show in Section~\ref{sec:results_group} results in a much more smooth evolution in $\tau$. 

We expect the constant of proportionality, $\alpha$, to be $\alpha \approx 1$, since we would expect accretion at a rate $\Gamma \approx 1$ for a time comparable to $t_\mathrm{c}$ to be disruptive enough to effectively reset the gravothermal evolution. In Section~\ref{sec:results_group}, we use $\alpha = 2.0$ to demonstrate the extended parametric model. This value is chosen retrospectively since, as highlighted later in Section~\ref{sec:results_group}, using the combination of parameters $(C = 0.75, \alpha = 2.0)$ in the extended model gives better results for predicting the density profile.

\section{Results: Group Simulation} \label{sec:results_group}

Having extracted halo samples from both the Group and Milky Way simulations, we now apply both the original and our extended parametric model to these samples, and assess the performance of each model in reproducing the properties of the simulated halos, specifically $V_\mathrm{max}$ at $z=0$, and the density profiles of $z=0$ halos. We begin by exploring how well the original parametric model performs on our sample of field halos from the Group simulation. Recall that the parametric model creates an SIDM halo density profile based on $V_{\max}$ and $R_{\max}$. Therefore, we assess how accurately this model predicts $V_{\max}$ and $R_{\max}$. 

\subsection{\texorpdfstring{$V_{\mathrm{max}}$}{Vmax} Comparisons}
Figure~\ref{fig:Vmax_examples} shows examples of the evolution of $V_\mathrm{max}$ with time for two SIDM halos in the Group simulation (purple lines), chosen to illustrate cases where the original parametric model performs well (left panel) and poorly (right panel). The dashed black lines in this figure show the evolution of $V_\mathrm{max}$ in the matched CDM halos, while the blue lines indicate the prediction from the original parametric model. While the original parametric model often performs well (as in the left panel), we find a significant fraction of cases where it fails to accurately match the evolution $V_\mathrm{max}$ measured from the simulation, usually by over-predicting the value of $V_{\max}$ as shown in the right panel.

How well either version of the parametric model performs can be measured by calculating the fractional error of $V_{\max}$ at $z=0$, $\delta_{V_\mathrm{max}}$---that is, the fractional offset between the predicted $V_{\max}$ at $z=0$ and that measured in the SIDM simulation
\begin{equation}
    \delta_{V_\mathrm{max}} = \frac{V_\mathrm{max, model}(z=0) - V^\mathrm{(SIDM)}_\mathrm{max}(z=0)}{V^\mathrm{(SIDM)}_\mathrm{max}(z=0)}.
    \label{eq:Vmax_percent_error}
\end{equation}
An ideal model would have  $\delta_{V_\mathrm{max}}$ clustered tightly around zero. 

\begin{figure}[H]
\centering
\includegraphics[width=0.9\linewidth]{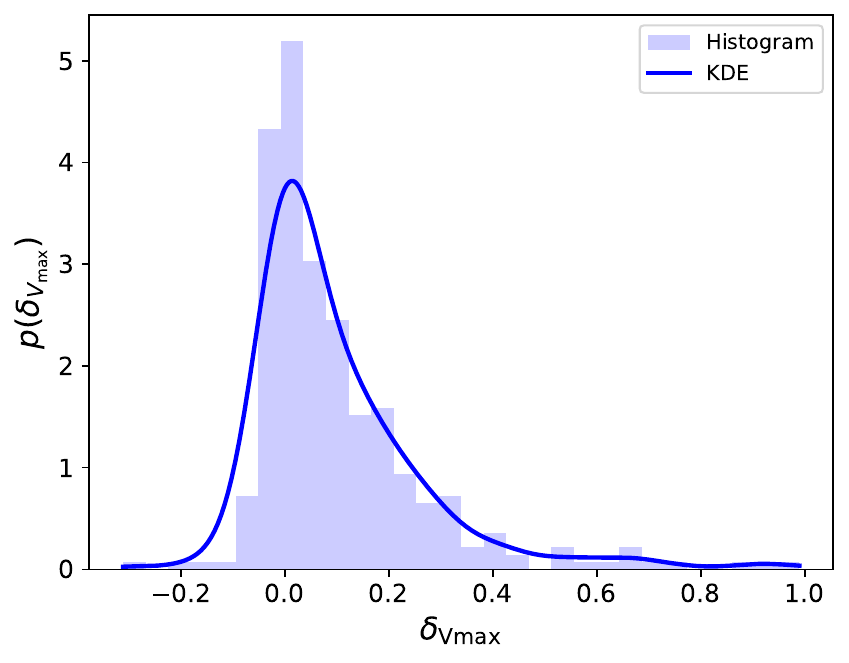}
\caption{The distribution of fractional offsets in $V_\mathrm{max}$ at $z=0$, $\delta_{V_\mathrm{max}}$, between the original parametric model and field halos selected from the Group simulation is shown by the blue histogram. The solid blue line shows a kernel density estimation (KDE) approximation to this distribution.}
\label{fig:hist_orig}
\end{figure}

Figure~\ref{fig:hist_orig} shows the distribution of $\delta_{V_\mathrm{max}}$ for the original parametric model for our sample of field halos in the Group simulation. It is clear that the original parametric model shows a peak around $\delta_{V_\mathrm{max}} = 0$---indicating accurate results for a large fraction of halos. However, there is also a substantial tail of the distribution to larger values $\delta_{V_\mathrm{max}}$. Notably, the distribution is highly skewed---the lowest negative values found are  $\delta_{V_\mathrm{max}} \approx -0.3$, while there are many halos for which $\delta_{V_\mathrm{max}} > 0.3$. In other words, for many halos, the original parametric model performs well in predicting $V_{\max}$, but when it fails, it tends to overpredict $V_{\max}$.  

\begin{figure}[H]
    \centering
    \includegraphics[width=0.9\linewidth]{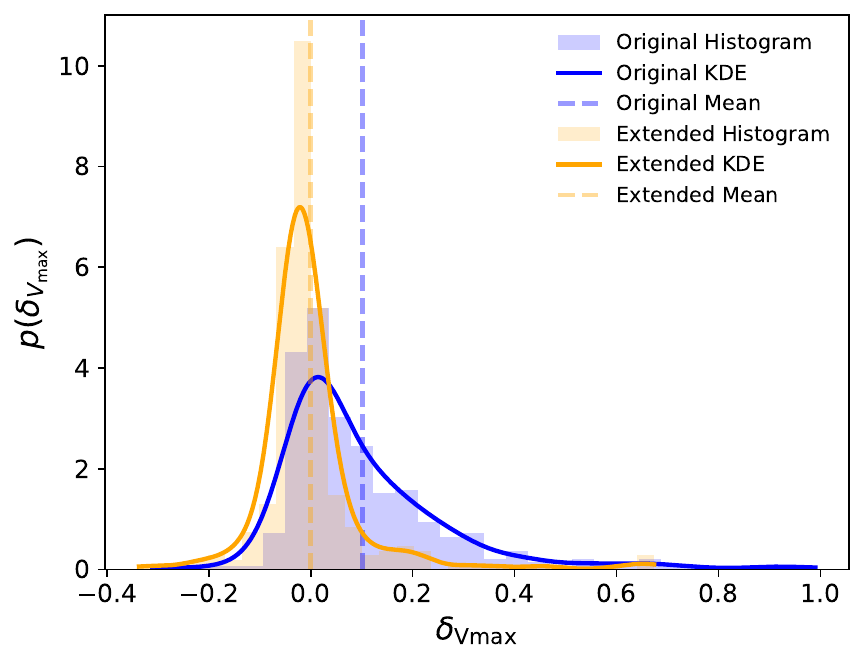}
    \caption{The distribution of fractional offsets in $V_\mathrm{max}$ at $z=0$, $\delta_{V_\mathrm{max}}$, between the extended parametric model (with $\alpha = 2.0$ and $C = 0.75$) and field halos selected from the Group simulation is shown by the orange histogram. The solid orange line shows a KDE approximation to this distribution. For comparison, the results of the original parametric model are shown in blue. Vertical dashed lines indicate the mean of each distribution, with the original model having a mean of $\langle \delta_{V_\mathrm{max}} \rangle = 0.10$ and the extended model having a mean of $\langle \delta_{V_\mathrm{max}} \rangle = 7.5 \times 10^{-4}$} 
    \label{fig:vmax_basic_comparison}
\end{figure}

Section~\ref{section:Mass accretion model} introduced an extension to the parametric model in which the parameter $\tau$ incorporates the effects of heating due to accretion of mass onto a halo. This extended model results in a slowing of the evolution of $\tau$ in the presence of significant mass accretion. As shown in section~\ref{section:Mass accretion model}, this new model introduces a parameter $\alpha$ that determines how strongly mass accretion slows the evolution of $\tau$. We will tune $\alpha$ to optimize the match between the predictions of the extended parametric model and N-body halos. We also choose to vary a second parameter, $C$, the constant appearing in Equation~(\ref{eq:tc}), which controls the overall normalization of $t_\mathrm{c}$, directly determining the overall timescale for core formation and core collapse. We note that the value of $C = 0.75$ has previously been determined by calibrating gravothermal models of SIDM halos to halos in N-body simulations \citep{Nishikawa_2020, yang2024parametric, Essig_2019}. Nevertheless, we allow this parameter to vary here and assess whether doing so leads to any improvement in our model. We are encouraged to do this based on findings from \cite{mace2025calibratingsidmgravothermalcatastrophe}, suggesting that $C$ is different for each halo based on its internal state and environment. However, neither $C$ nor $\alpha$ will change on a per-halo basis; instead, we choose values of $C$ and $\alpha$ and apply our extended model given those values for all halos in our sample.

\begin{figure*}[t]
\begin{center}
\begin{tabular}{cc}
    \includegraphics[width=0.45\linewidth]{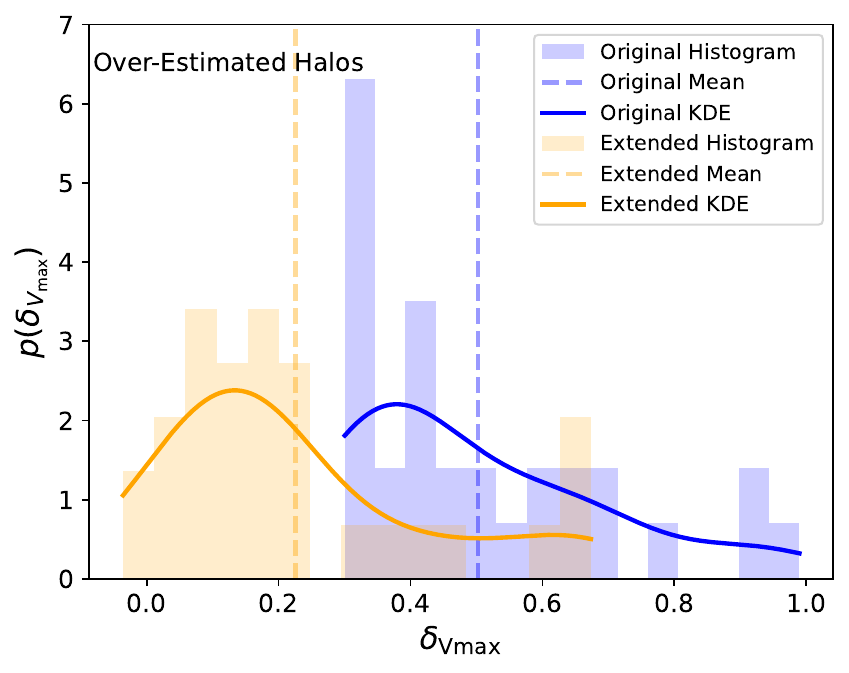} &
    \includegraphics[width=0.45\linewidth]{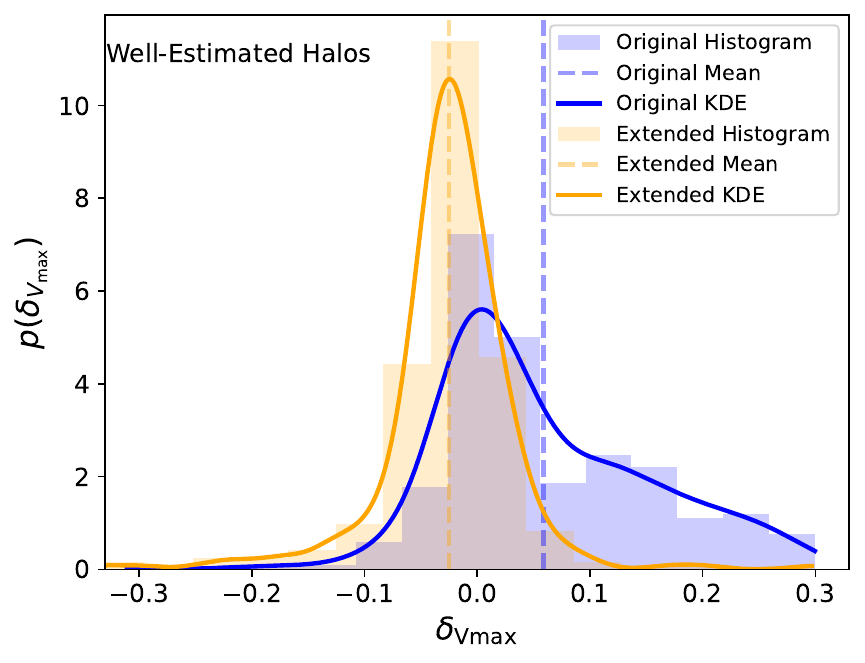}
\end{tabular}
\end{center}

\caption{The left panel shows halos for which the original parametric model predicted $\delta_{V_\mathrm{max}} > 0.3$, while the right panel shows those for which it predicted $\delta_{V_\mathrm{max}} < 0.3$. The distribution of $\delta_{V_\mathrm{max}}$ for the original model and extended model is shown for both samples of halos to highlight how differently the predictions for $\delta_{V_\mathrm{max}}$ change with the extended model for halos that are over-estimated vs well-estimated in the original model.}
\label{fig:comparisons}
\end{figure*}
The extended parametric model is applied to the same population of field halos as was used for the original parametric model. Figure~\ref{fig:vmax_basic_comparison} shows the results for the extended parametric model for the parameters $C = 0.75$ and $\alpha = 2.0$, where the value of $\alpha = 2.0$ is chosen arbitrarily to illustrate how the extended model behaves compared to the original model. Figure \ref{fig:vmax_basic_comparison} shows the resulting distribution of $\delta_{V_\mathrm{max}}$ for the extended model, clearly showing that the distribution of $\delta_{V_\mathrm{max}}$ becomes tighter and more centered on zero compared to the original model. 

Figure~\ref{fig:comparisons} further illustrates the behavior of our extended model. In this figure, we show the same distribution functions, but we split the sample of halos into those for which the original parametric model predicted $\delta_{V_\mathrm{max}} > 0.3$ (left panel), and those for which it predicted $\delta_{V_\mathrm{max}} < 0.3$ (right panel). It can be seen that the greatest effect of the extended model is on halos for which $V_\mathrm{max}$ was substantially overestimated by the original model (left panel), reducing the mode of the distribution of $\delta_{V_\mathrm{max}}$ from around 0.4 to 0.15. The extended model also affects halos for which the original model performed better, $\delta_{V_\mathrm{max}} < 0.3$ (right panel), reducing the mode of the distribution from around 0.0 to $-0.025$, and predicting more negative values of $\delta_{V_\mathrm{max}}$ for some halos than did the original model.

\subsection{\texorpdfstring{$R_{\mathrm{max}}$}{Rmax} Comparisons}
\begin{figure}
\centering
\includegraphics[width=0.9\linewidth]{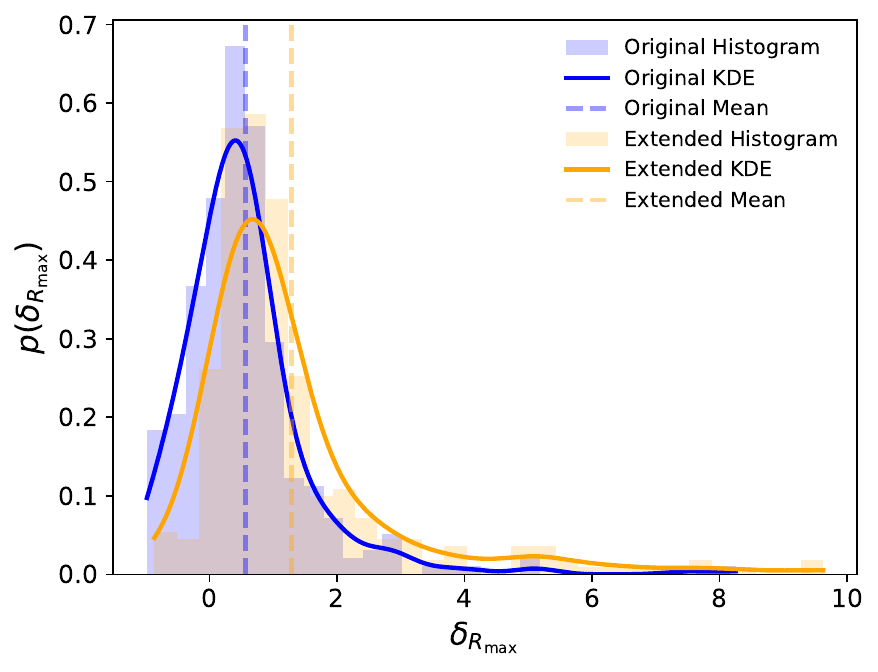}
\caption{The distribution of fractional offsets in $R_\mathrm{max}$ at $z=0$, $\delta_{R_\mathrm{max}}$, between the original parametric model and field halos selected from the Group simulation is shown by the blue histogram. The same is shown for the extended parametric model in orange. KDE of each histogram is shown to better illustrate the spread of $\delta_{R_\mathrm{max}}$ values. Note that the mean of $\delta_{R_\mathrm{max}}$, $\langle \delta_{R_\mathrm{max}}\rangle$ for the original model is $\langle \delta_{R_\mathrm{max}}\rangle = 0.57$. For the extended model the mean is $\langle \delta_{R_\mathrm{max}}\rangle = 1.29$.} 
\label{fig:Rmax_basic_distr_group}
\end{figure}

Figure~\ref{fig:Rmax_basic_distr_group} presents $\delta_{R_\mathrm{max}}$ for the original model and the extended model for the Group simulations, where $\delta_{R_\mathrm{max}}$ is defined similarly to $\delta_{V_\mathrm{max}}$ as:

\begin{equation}
    \delta_{R_\mathrm{max}} = \frac{R_\mathrm{max, model}(z=0) - R^\mathrm{(SIDM)}_\mathrm{max}(z=0)}{R^\mathrm{(SIDM)}_\mathrm{max}(z=0)}.
    \label{eq:Rmax_percent_error}
\end{equation}

The extended parametric model does not show improvement for $R_{\mathrm{max}}$ since the distribution of $\delta_{R_{\max}}$ for the extended model in Figure~\ref{fig:Rmax_basic_distr_group} has a higher mean. From the distribution, it is evident that the extended parametric model tends to over-predict $R_{\mathrm{max}}$ more than the original parametric model. Specifically, for the original model $\langle \delta_{R_\mathrm{max}} \rangle = 0.57$ with a standard deviation of $\mathrm{std}(\delta_{R_\mathrm{max}}) = 1.06$, while the extended model has $\langle \delta_{R_\mathrm{max}} \rangle = 1.29$ with $\mathrm{std}(\delta_{R_\mathrm{max}}) = 1.62$. Although the peaks of the distributions are much closer, the extended model has more halos for which the $\delta_{R_\mathrm{max}}$ is large. This is because the extended model keeps $\tau$ at lower values than the original model, which works well for predicting $V_\mathrm{max}$ but tends to delay how fast the core shrinks in the extended model, thus over-predicting $R_\mathrm{max}$. The reason why our extended model does a much better job at predicting $V_\mathrm{max}$ is also the reason why it struggles to predict $R_\mathrm{max}$. 

\begin{figure}
\begin{center}
    \includegraphics[width=0.9\linewidth]{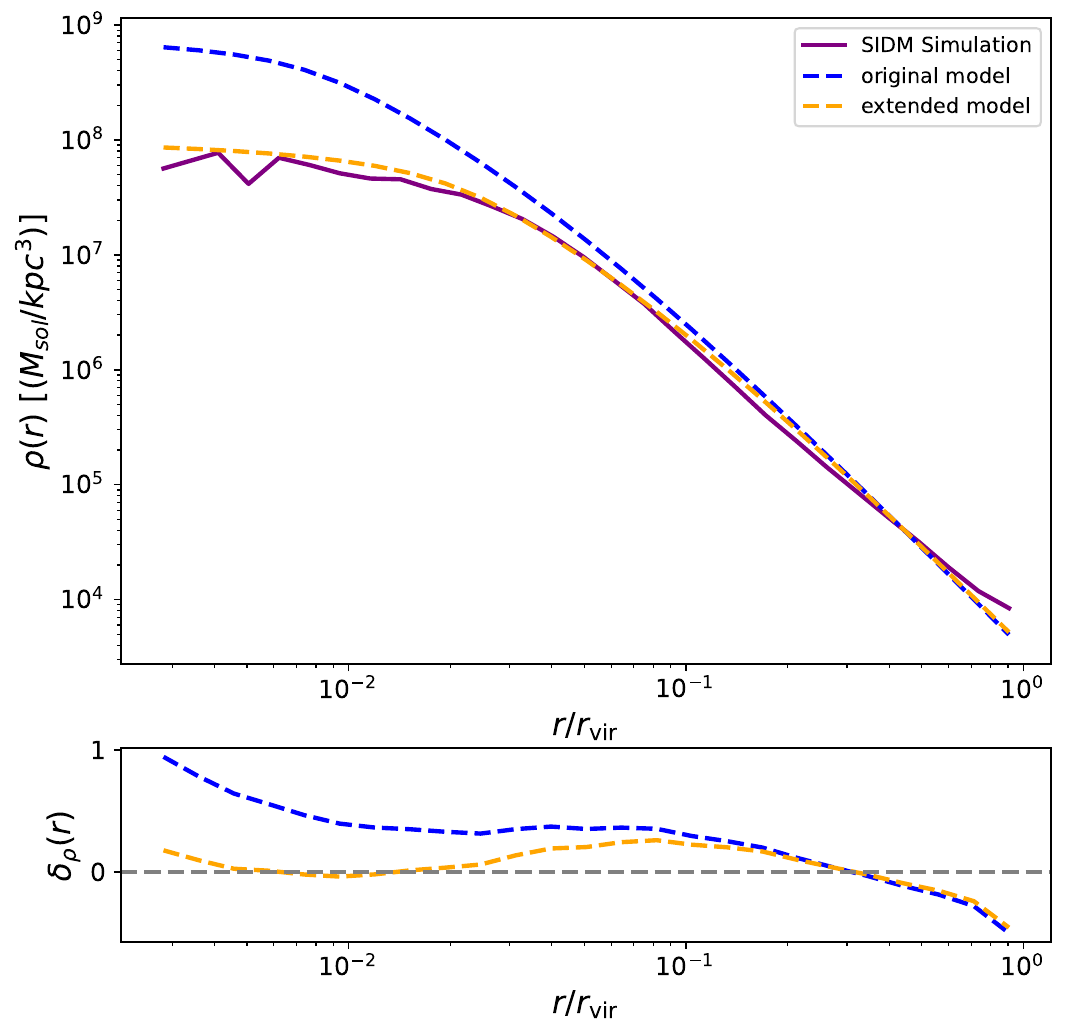}
\end{center}
\caption{The density profile of a halo selected from the group simulation for which the original parametric model fails significantly. In the upper panel, the black solid line represents the measured N-body profile, while the orange and green dot-dashed lines represent predictions from the original and extended parametric models, respectively. The lower panel shows the error in the models' profile based on the SIDM simulation, where the error is defined in Equation~\ref{eq:delta_rho}.}
\label{fig:profile_good}
\end{figure}

For the extended model, the improvements in the distribution of $\delta_{V_\mathrm{max}}$ in Figure~\ref{fig:vmax_basic_comparison} are more considerable than the slightly more skewed distribution of $\delta_{R_\mathrm{max}}$ in Figure~\ref{fig:Rmax_basic_distr_group}. The top panel of Figure~\ref{fig:profile_good} shows an example for a halo selected to have $\delta_{V_\mathrm{max}} > 0.3$ with the original parametric model. It can be seen that this inaccuracy in the predicted $V_\mathrm{max}$ corresponds to a substantial overestimate of the core density---by a factor of around eight at small radii. The extended parametric model performs significantly better in matching the N-body density profile across the entire range of radii shown, as can be seen in the lower panel, where we display the fractional difference between the predictions of the parametric models and the N-body simulation.


\begin{figure*}[t]
\begin{center}
\begin{tabular}{cc}
    \includegraphics[width=0.9\linewidth]{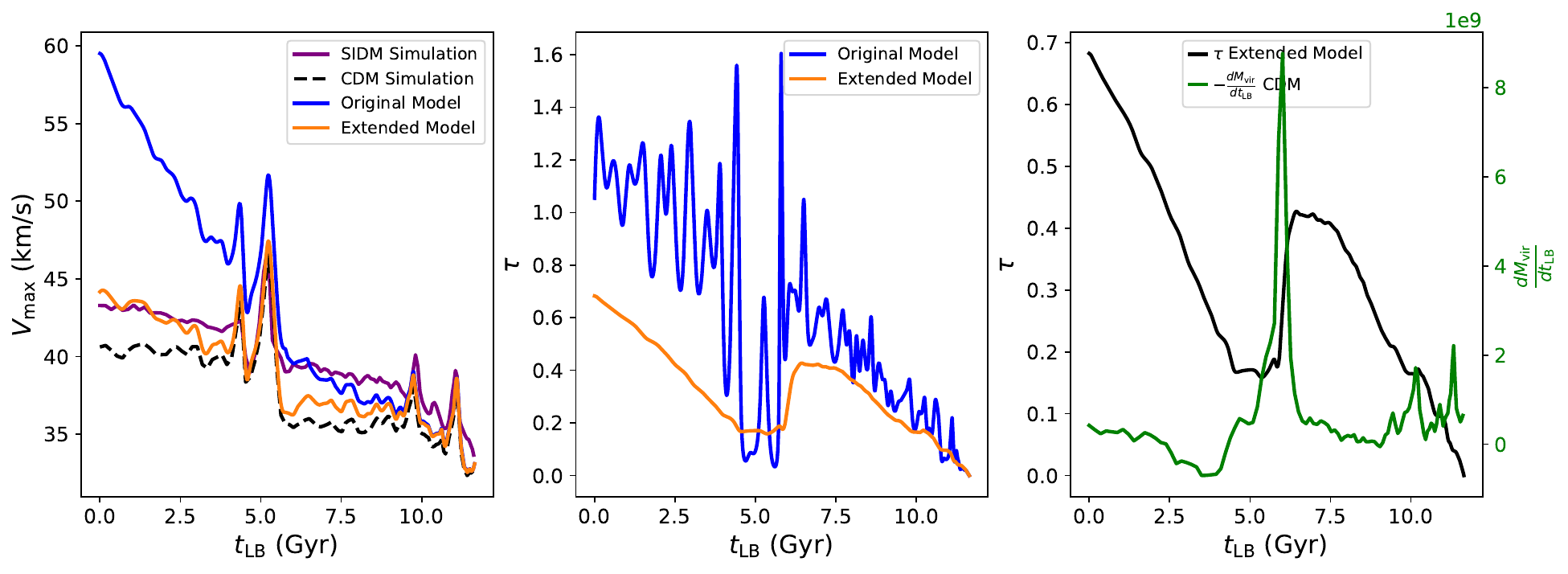} 
\end{tabular}
\end{center}
\caption{The left panel shows $V_\mathrm{max}$ values predicted by both original and extended models, and also those obtained from the N-body simulations. The center panel shows $\tau$ for the extended and original model, while the right panel shows $\tau$ for the extended model against the mass accretion rate to illustrate how our extended model drives $\tau$ to lower values during periods of rapid growth.}
\label{fig:Vmax_example1}
\end{figure*}

\begin{figure*}[t]
\begin{center}
\begin{tabular}{cc}
    \includegraphics[width=0.9\linewidth]{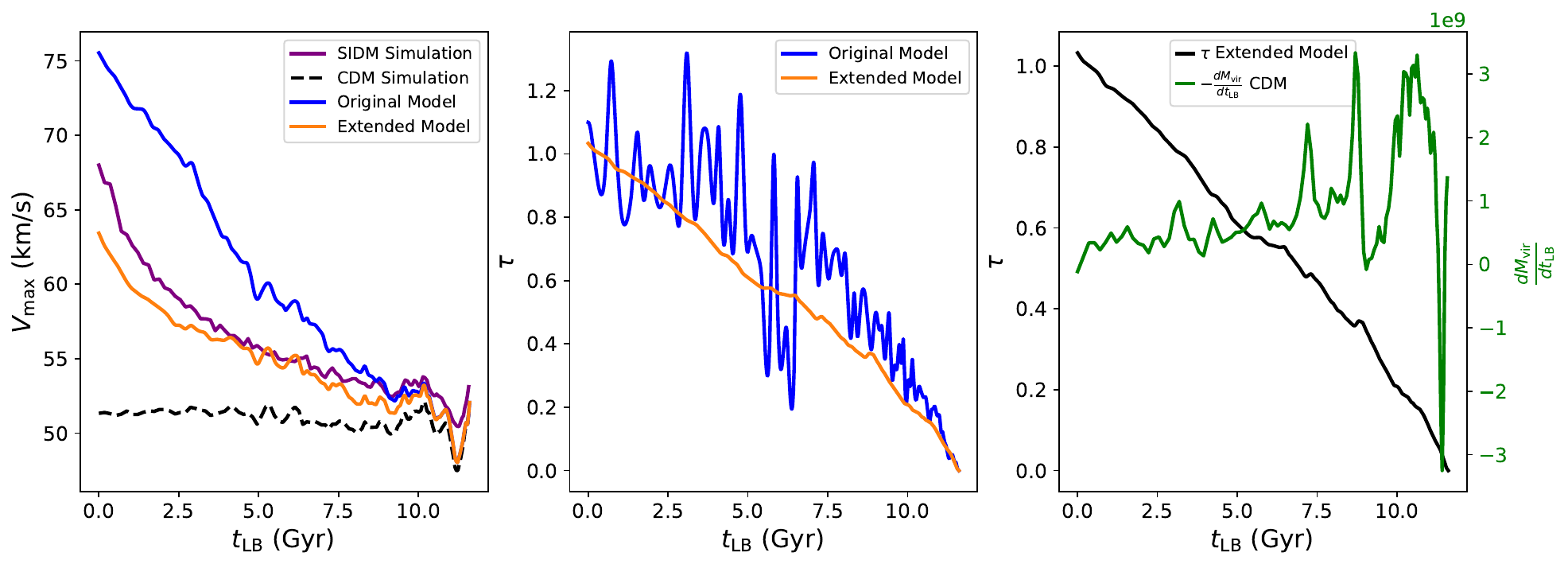} 
\end{tabular}
\end{center}

\caption{As Figure~\ref{fig:Vmax_example1}, but for a different pair of CDM and SIDM halo from the Group simulation. This example highlights how the extended model can also better predict $V_\mathrm{max}$ values for halos that experience more steady mass accretion.}
\label{fig:Vmax_example2}
\end{figure*}





Figure~\ref{fig:Vmax_example1} and Figure~\ref{fig:Vmax_example2} illustrate how our extended model is able to better predict $V_\mathrm{max}$. Halos that benefit the most from our extended model tend to experience either short periods of fast mass accretion, such as a merger, or significant mass growth for an extended period of time. The left panels of Figure~\ref{fig:Vmax_example1} and Figure~\ref{fig:Vmax_example2} show $V_\mathrm{max}$ predicted by the original and extended model against the $V_\mathrm{max}$ obtained from the paired CDM and SIDM halos from the N-body simulations. The center panel shows the $\tau$ of each model versus lookback time, $t_\mathrm{LB}$. The right panel shows the $\tau$ of the extended model and the rate of mass accretion, $- \mathrm{d}M_\mathrm{vir}/\mathrm{d}t_\mathrm{LB}$ (calculated using the data for $M_\mathrm{vir}$ from \textsc{Rockstar} and \textsc{Consistent-Trees}).

Figure~\ref{fig:Vmax_example1} is an example of paired halos from the Group simulation that go through an event that results in back-to-back spikes in $V_\mathrm{max}$ at $t_\mathrm{LB} \approx 4$--$6~\mathrm{Gyrs}$ as seen in the left panel. These spikes in $V_\mathrm{max}$ coincide with the large spike in mass accretion rate, $- \mathrm{d}M_\mathrm{vir}/\mathrm{d}t_\mathrm{LB}$, seen in the right panel in green, suggesting a merger event is the cause. The original model (blue) begins to over-predict $V_\mathrm{max}$ after the two spikes, while our extended model (orange) manages to stay closer to the $V_\mathrm{max}$ from the SIDM N-body simulation (purple). The center panel demonstrates why $V_\mathrm{max}$ increases rapidly after the spikes for the original model. The original model has large spikes in $\tau$ and remains at high values of $\tau \approx 1$. Recall that the third term on the right-hand side of Equation~\ref{eq:vmaxintegral} is the one responsible for increasing $V_\mathrm{max}$ as a function of $\tau$ in order to mimic the effects of SIDM. Looking at Equation~\ref{eq:param_Vmax_cal}, it is evident that higher values of $\tau$---especially ones close to unity---will result in the third term of Equation~\ref{eq:vmaxintegral} rapidly increasing $V_\mathrm{max}$ as if the SIDM halo is in core collapse. In comparison, our extended model manages to keep $\tau$ at lower values during and after the spikes in $V_\mathrm{max}$ as seen in the center panel. In the right panel, we see that $\tau$ for the extended model drops down to lower values just as our CDM halo experiences rapid mass accretion. Bringing $\tau$ back down and keeping it at lower values after the event results in the extended model predicting the $V^\mathrm{(SIDM)}_\mathrm{max}$ more closely.

\begin{figure*}[t]
\begin{center}
\begin{tabular}{cc}
    \includegraphics[width=0.45\linewidth]{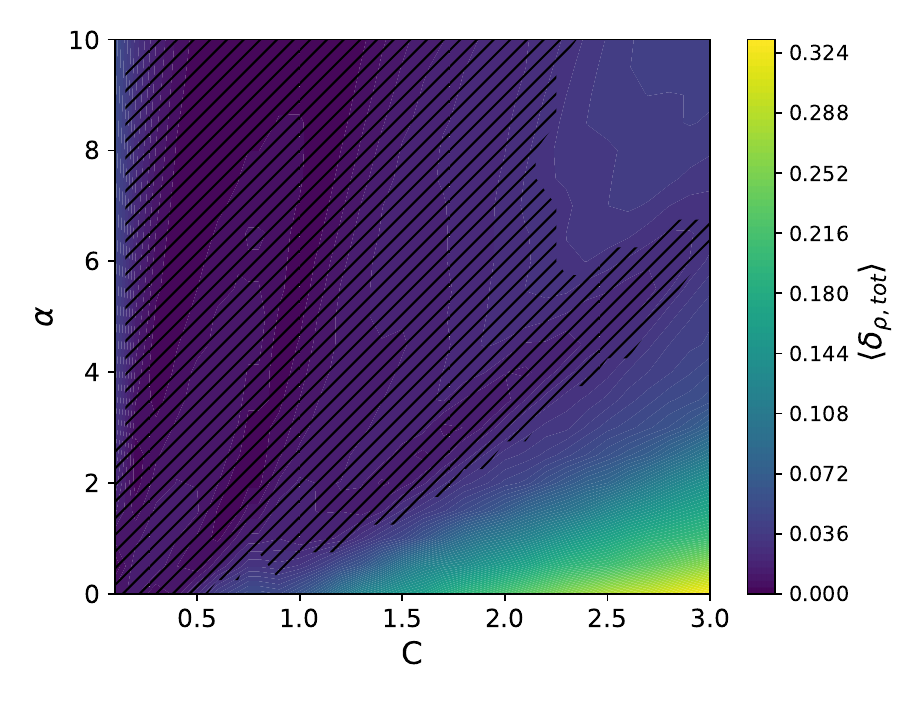}  &  \includegraphics[width=0.45\linewidth]{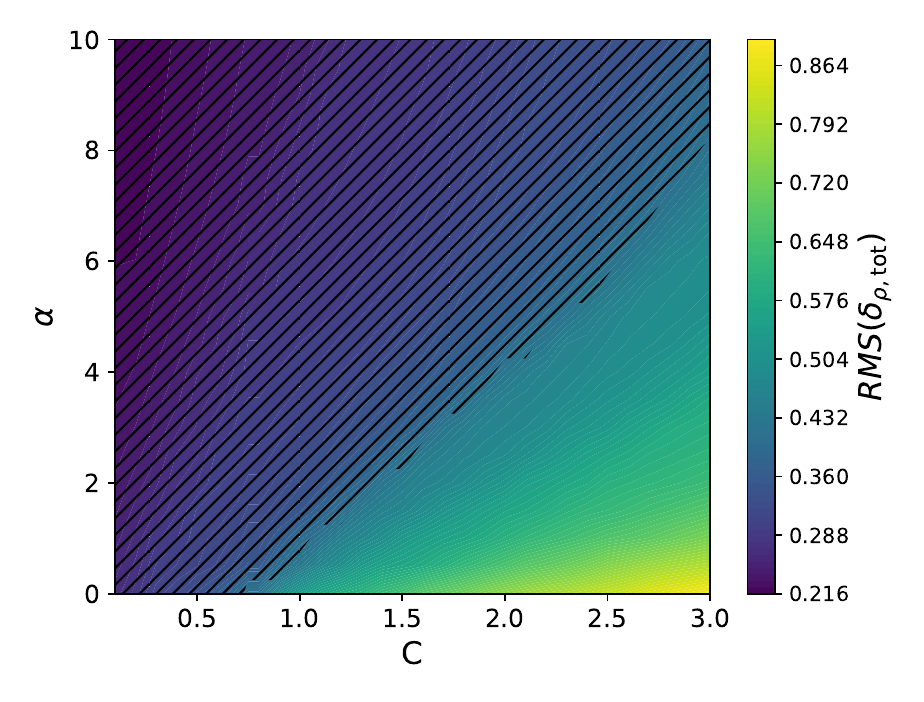}
\end{tabular}
\end{center}
\caption{2D color maps measuring the performance of the extended parametric model in matching density profiles of SIDM halos in the Group simulation. The left panel shows $\langle \delta_{\rho,\mathrm{tot}}\rangle$ for a range of $C$ and $\alpha$ for the extended parametric model, while the right panel shows $\mathrm{RMS}(\delta_{\rho,\mathrm{tot}})$. The hatched area depicts the region of $(C, \alpha)$ parameters for which the extended parametric model performs better than the original parametric model.}
\label{fig:profile_parameter_stats_group_v1}
\end{figure*}

Figure~\ref{fig:Vmax_example2} illustrates another example of paired halos from the Group simulation, but this time the halo simply experiences steady mass accretion for $t_\mathrm{LB} < 7~\mathrm{Gyr}$ as seen in the right-most panel of the figure. As a result, there is no significant dip in $\tau$ for our extended model that coincides with a fast mass accretion event, yet the left-most panel shows that our extended model performs better than the original model. As soon as $t_\mathrm{LB} \approx 8~\mathrm{Gyr}$, the original model begins to over predict $V_\mathrm{max}$. The center panel of the figure shows the original model to have larger values for $\tau$ in general and also frequent spikes. Over time, the higher values for $\tau$, especially with frequent spikes to larger values, will result in the parametric model increasing $V_\mathrm{max}$ faster in comparison to our extended model. Both examples in Figure~\ref{fig:Vmax_example1} and Figure~\ref{fig:Vmax_example2} illustrate how the extended model's choice to calculate $\tau$ with Equation~\ref{eq:tau_dot} results in more accurate predictions for $V_\mathrm{max}$. 

\subsection{Calibrating \texorpdfstring{$C$}{C} and \texorpdfstring{$\alpha$}{alpha}}
We can further assess differences between the original and extended parametric models by examining the predicted density profiles of SIDM halos, following the approach described in Section~\ref{section:param_model}, and comparing them to those measured directly from the Group simulation. This is a stronger test than simply considering $V_\mathrm{max}$ (since $V_\mathrm{max}$ is fully-determined by the density profile, while the reverse is not true). We also use the comparison of the predicted profiles to the N-body simulations to determine which $C$ and $\alpha$ values are the optimal choice. 

To determine optimal choices for the parameters $C$ and $\alpha$ of our extended parametric model, we compute predicted density profiles for all halos in our Group sample at $z=0$. Recall that this is done by taking the $\rho_\mathrm{s}$, $r_\mathrm{s}$, and $r_\mathrm{c}$ as predicted by the parametric model at $z=0$ and using Equation~\ref{eq:sidm_profile} to obtain the density profile which we will call $\rho_\mathrm{model}(r)$. We then bin data from the SIDM N-body simulation to get the actual density profile from the simulation, which we will refer to as $\rho_\mathrm{data}(r)$. 
With $\rho_\mathrm{data}(r)$ and $\rho_\mathrm{model}(r)$ we are able to calculate

\begin{equation}
    \delta_\rho(r) = 2  \frac{\rho_\mathrm{model}(r) - \rho_\mathrm{data}(r)}{\rho_\mathrm{model}(r) + \rho_\mathrm{data}(r)}.
    \label{eq:delta_rho}
\end{equation}

We calculate $\delta_\rho(r)$ for all SIDM halos on a range of $0.02R_\mathrm{vir}<r<0.9R_\mathrm{vir}$ for each halo. To assess which set of parameter choices for $C$ and $\alpha$ makes the extended model outperform the original model when it comes to predicting halo density profiles, we look at the mean and root-mean-squared (RMS) of the $\delta_\rho(r)$ function. For each halo, we calculate the mean of $ \delta_\rho(r)$ over the range of $0.02R_\mathrm{vir}<r<0.9R_\mathrm{vir}$, which we will call $\delta_{\rho,\mathrm{tot}}$. We then find the average and RMS of $\delta_{\rho,\mathrm{tot}}$ over all halos in our sample of field halos, giving us $\langle \delta_{\rho,\mathrm{tot}}\rangle$ and $\mathrm{RMS}(\delta_{\rho,\mathrm{tot}})$. 

\begin{figure}[b]
\begin{center}
    \includegraphics[width=0.9\linewidth]{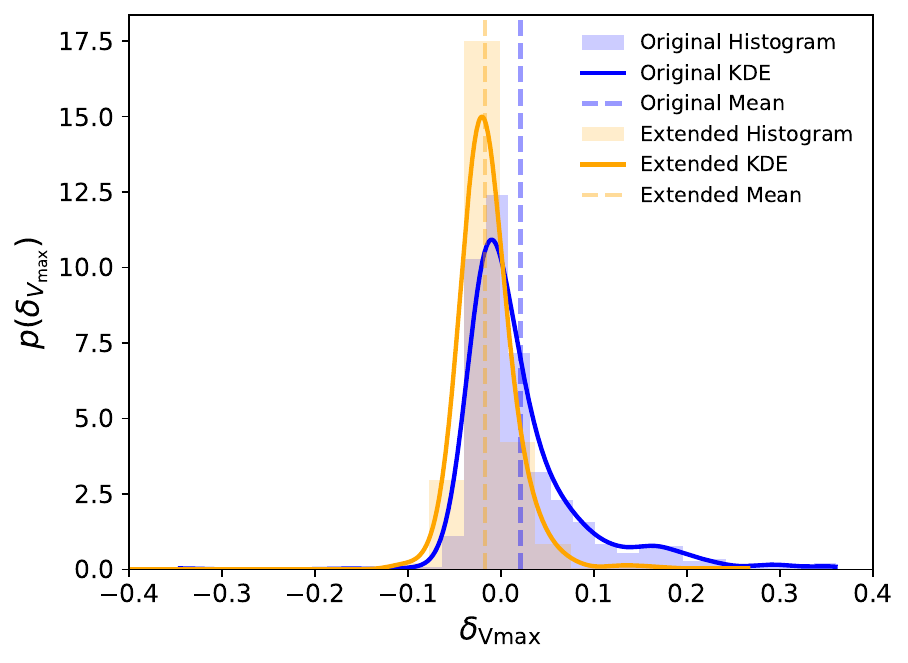}
\end{center}
\caption{The distribution of $\delta_{V_\mathrm{max}}$ for the original model and extended model for the sample of halos in the Milky Way simulations. The original model has a mean of $\langle \delta_{V_\mathrm{max}} \rangle = 0.021$ and the extended model has a mean of $\langle \delta_{V_\mathrm{max}} \rangle = -0.017$}
\label{fig:vmax_basic_comparison_vd}
\end{figure}

Figure~\ref{fig:profile_parameter_stats_group_v1} shows the values for $\langle \delta_{\rho,\mathrm{tot}}\rangle$ in the left panel and $\mathrm{RMS}(\delta_{\rho,\mathrm{tot}})$ in the right panel for different values of $C$ and $\alpha$. The hatched regions represent the combinations of $(C, \alpha)$ for which the extended model outperforms the original model by having lower $\langle \delta_{\rho,\mathrm{tot}}\rangle$ or $\mathrm{RMS}(\delta_{\rho,\mathrm{tot}})$ values. It can be seen that for $C<2.0$, the extended model outperforms the original model for most $\alpha$ values. However, there are clear ``valleys'' in the 2D colormap where $\langle \delta_{\rho,\mathrm{tot}}\rangle$ is clearly somewhat lower.


\begin{figure*}[t]
\begin{center}
\begin{tabular}{cc}
    \includegraphics[width=0.45\linewidth]{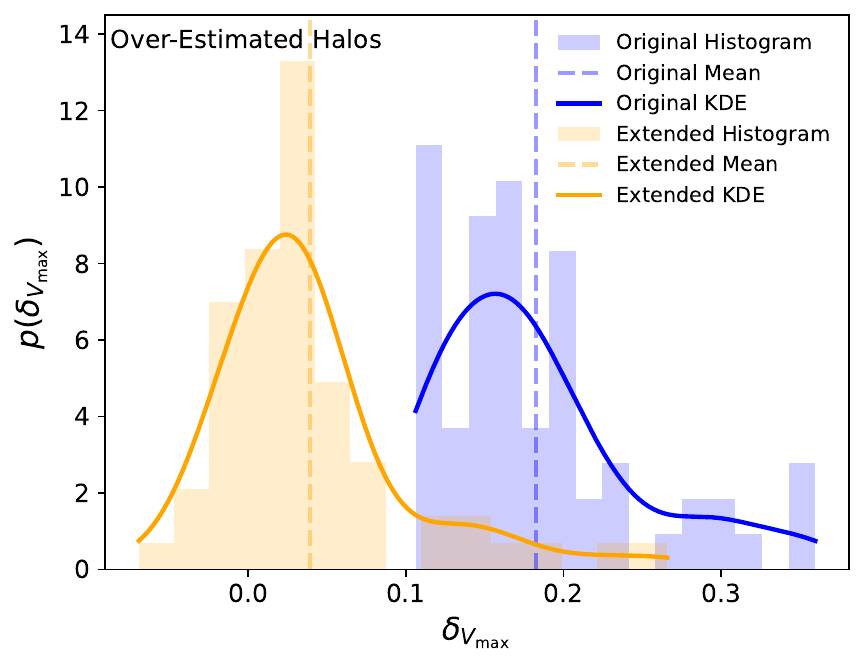} &   \includegraphics[width=0.45\linewidth]{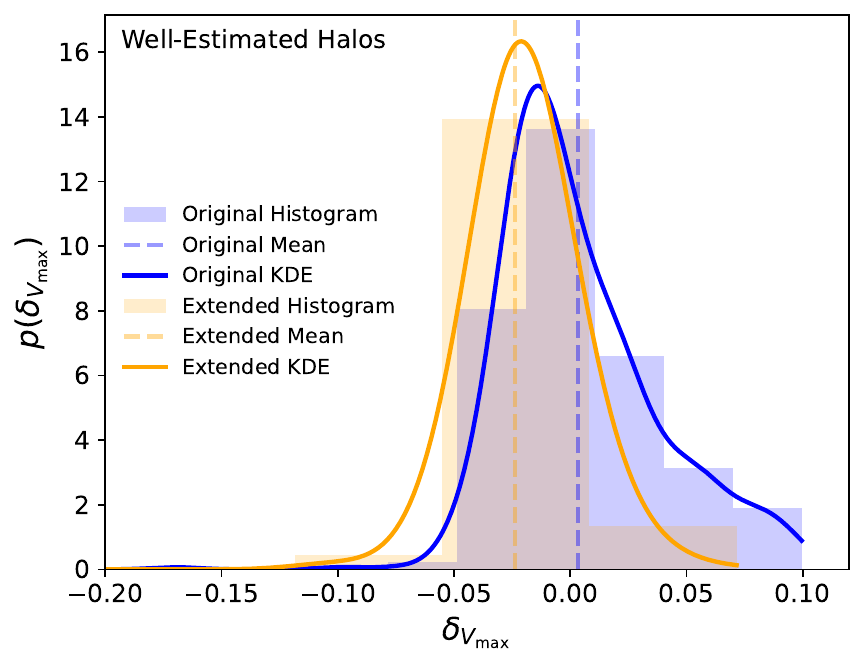}
\end{tabular}
\end{center}
\caption{The distribution of fractional offsets in $V_\mathrm{max}$ at $z=0$, $\delta_{V_\mathrm{max}}$, between the extended parametric model (with $\alpha = 2.0$ and $C = 0.75$) and field halos selected from the Milky Way simulation is shown by the orange histogram. The solid orange line shows a KDE approximation to this distribution. For comparison, the results of the original parametric model are shown in blue. Vertical dashed lines indicate the mean of each distribution. The left panel shows halos for which the original parametric model predicted $\delta_{V_\mathrm{max}} > 0.1$, while the right panel shows those for which $\delta_{V_\mathrm{max}} < 0.1$ is predicted.}
\label{fig:comparisons_vd}
\end{figure*}

Note that on a per halo basis, the best set of parameters to predict the SIDM profile and $V_\mathrm{max}$ does vary, but Figure~\ref{fig:profile_parameter_stats_group_v1} illustrates how well each combination of parameters, $(C, \alpha)$, predicts SIDM profiles across all halos in our Group sample. It is also reassuring that the commonly used values for $C$ in the literature, such as $C = 0.75$, outperform the original model for a range of $\alpha$ values according to Figure~\ref{fig:profile_parameter_stats_group_v1}. 

An interesting feature to note in Figure~\ref{fig:profile_parameter_stats_group_v1} is the presence of three shallow valleys in the left panel that plots $\langle \delta_{\rho,\mathrm{tot}}\rangle$. These darker regions highlight $(C, \alpha)$ values that have $\langle \delta_{\rho,\mathrm{tot}}\rangle$ slightly lower than the rest. That is, there seem to be three different clear sets of $(C,\alpha)$ values for which our extended model performs better. We speculate that this may be due to there being distinct subsets of halo merger histories in our sample of field halos for which a particular set of $(C, \alpha)$ values is strongly favored. Work by \cite{mace2025calibratingsidmgravothermalcatastrophe} already suggests that parameters such as $C$ (which they refer to as $\beta$) are unique to individual halos. We leave the work of exploring unique values of $(C, \alpha)$ for each individual halo for a future paper. However, if we are only interested in looking for a singular best $(C,\alpha)$ value to fit all of the halos in our Group sample, it is clear that for any given $C$ value---that can be calibrated externally, such as with N-body simulations---we can find the best $\alpha$ using Figure~\ref{fig:profile_parameter_stats_group_v1}. 


\section{Results: Milky Way simulations} \label{sec:results_MW}

To assess how well the extended parametric model works in a different range of halo masses and a different set of parameters, $\sigma_0$ and $w$, for our velocity dependent SIDM model, we repeat the above analysis for our Milky Way simulation pair and present the results below.

Figure~\ref{fig:vmax_basic_comparison_vd} shows the distribution of $\delta_{V_\mathrm{max}}$ for halos from the Milky Way simulations for the original and extended model. Once again, for the original model, the distribution is skewed to higher values of $\delta_{V_\mathrm{max}}$, suggesting a similar trend of over-prediction present in the Group simulations. For consistency, we choose the same parameter values of $C = 0.75$ and $\alpha = 2.0$ for the extended model shown in Figures~\ref{fig:vmax_basic_comparison_vd} and \ref{fig:comparisons_vd}. As in the Group simulations, we find here that the extended parametric model results in a distribution of $\delta_{V_\mathrm{max}}$ that is more tightly peaked around zero and less skewed. In this case, the mean of the distribution is not significantly closer to zero than in the original model, merely being offset to negative values, but by a comparable absolute amount.

\begin{figure}
\begin{center}
\includegraphics[width=0.9\linewidth]{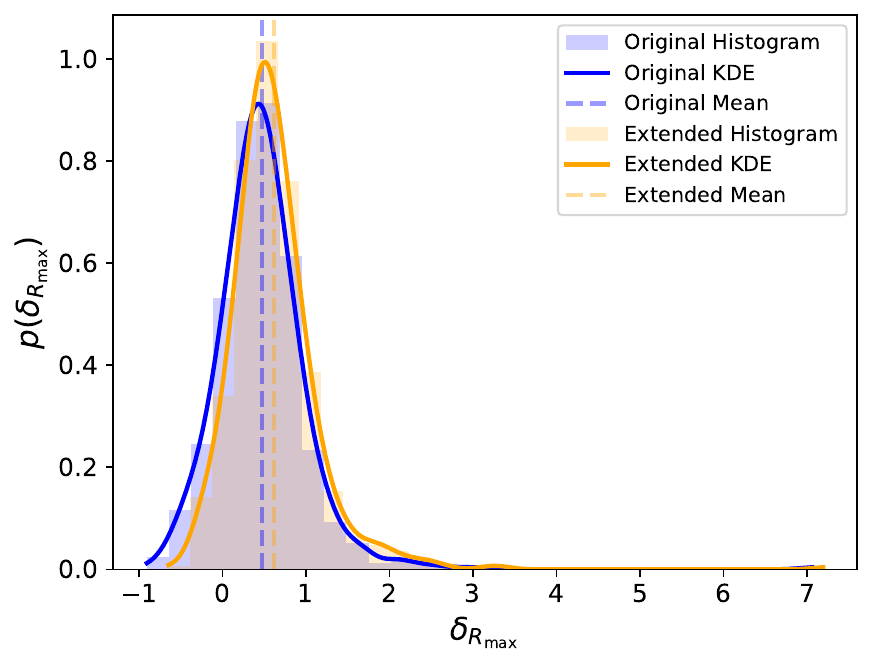}
\end{center}
\caption{The distribution of $\delta_{R_\mathrm{max}}$ resulting from the original parametric model applied to halos from the Milky Way simulations is chosen by the blue histogram, with the blue line showing a kernel density estimate of this distribution function. The original model has a mean of $\langle \delta_{R_\mathrm{max}} \rangle = 0.48$ and the extended model has a mean of $\langle \delta_{R_\mathrm{max}} \rangle = 0.62$}
\label{fig:Rmax_basic_distr_final}
\end{figure}

\begin{figure*}
\begin{center}
\begin{tabular}{cc}
    \includegraphics[width=0.45\linewidth]{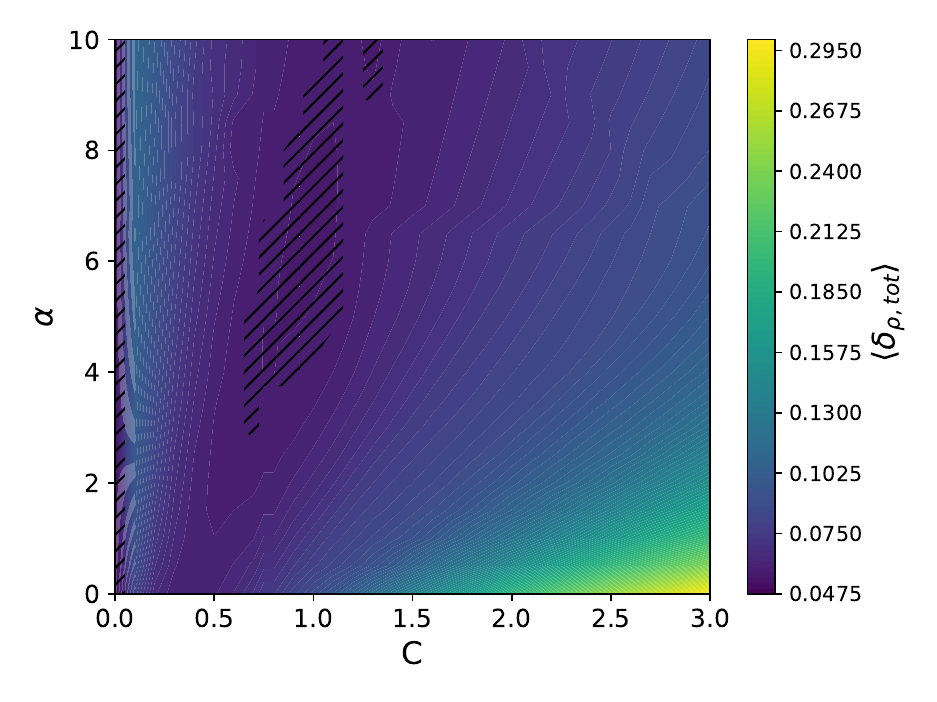}  &  \includegraphics[width=0.45\linewidth]{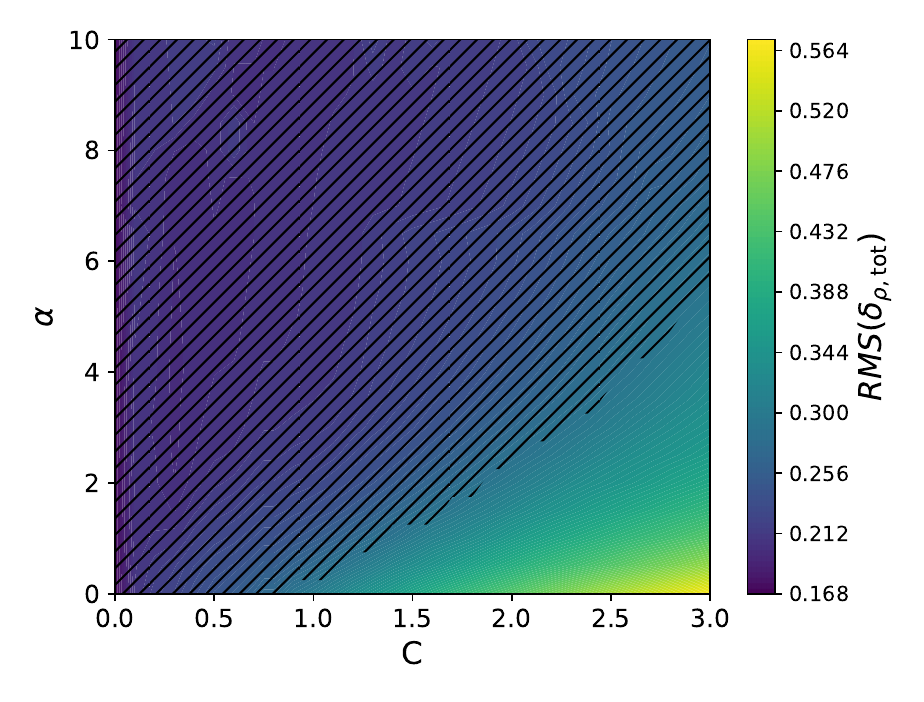}
\end{tabular}
\end{center}
\caption{2D colormaps measuring the performance of the extended parametric model in matching density profiles of SIDM halos in the Milky Way simulation. The left panel shows $\langle \delta_{\rho,\mathrm{tot}}\rangle$ as a function of $C$ and $\alpha$ for the extended parametric model, while the right panel shows $\mathrm{RMS}(\delta_{\rho,\mathrm{tot}})$. The hatched area depicts the region of $(C, \alpha)$ parameters for which the extended parametric model performs better than the original parametric model.}
\label{fig:profile_parameter_stats_final_v1}
\end{figure*}

We can once again separate our sample of halos from the Milky Way simulation into subsamples for which $\delta_{V_\mathrm{max}}$ is well-estimated by the original parametric model ($\delta_{V_\mathrm{max}} < 0.1$), and those that are over-estimated ($\delta_{V_\mathrm{max}} > 0.1$). Figure~\ref{fig:comparisons_vd} shows these two subsamples, and demonstrates that the improvement in the extended parametric model arises mostly in the over-estimated subsample, while keeping the distribution of $\delta_{V_\mathrm{max}}$ of well-estimated halos still centered around zero. This is similar to our finding for the Group simulation. The subsample of over-estimated halos performs much better with the extended model versus the original. On the other hand, the subsample of well-estimated halos performs on average a little worse since  $\langle\delta_{V_\mathrm{max}}\rangle$ for the extended model is a negative value of $\langle\delta_{V_\mathrm{max}}\rangle = -0.024$. In contrast, the original model has $\langle\delta_{V_\mathrm{max}}\rangle = 0.0033$ for well-estimated halos. Nevertheless, the mean $\langle \delta_{V_{\max}}\rangle$ shifts only by around $0.027$ while the improvement in the over-estimated halos is much larger at around $0.1$. As such, the extended parametric model still provides a more accurate overall prediction of $V_{\max}$.

Figure~\ref{fig:Rmax_basic_distr_final} compares the distribution for $\delta_{R_{\max}}$ for the Milky Way simulation for the original versus the extended parametric model. In the Milky Way simulation, both the original and extended parametric models perform similarly, having $\delta_{R_{\max}}$ histograms that largely overlap with each other. Specifically, for the original model $\langle \delta_{R_{\max}} \rangle = 0.48$ with a standard deviation of $\mathrm{std}(\delta_{R_{\max}}) = 0.56$, while the extended model has a mean of $\langle \delta_{R_{\max}} \rangle = 0.62$ with a standard deviation of $\mathrm{std}(\delta_{R_{\max}}) = 0.55$. For the Milky Way simulation the extended model performs quite similar to the original model when it comes to predicting $R_{\max}$ across the sample of Milky Way field halos. Unlike in the Group simulation, the extended model predicts $V_{\max}$ better at little to no cost to its ability to predict $R_{\max}$ in the Milky Way simulation. 

Finally, we perform the same analysis on the density profile of the Milky Way halos as we did on the Group halos. Just as before, we show $\langle \delta_{\rho,\mathrm{tot}}\rangle$ on the left panel and $\mathrm{RMS}(\delta_{\rho,\mathrm{tot}})$ on the right panel of Figure~\ref{fig:profile_parameter_stats_final_v1}. The right panel in Figure~\ref{fig:profile_parameter_stats_final_v1} is similar to that in Figure~\ref{fig:profile_parameter_stats_group_v1} in showing that there are many parameter combinations of $(C, \alpha)$ that have a better RMS density errors than the original model. However, when it comes to the  $\langle \delta_{\rho,\mathrm{tot}}\rangle$ there are many fewer parameter options for the Milky Way subsample of halos that perform better than the original model. In fact, the left panel of Figure~\ref{fig:profile_parameter_stats_final_v1} shows a better average $\langle \delta_{\rho,\mathrm{tot}}\rangle$ for the extended model for only a handful of parameter combinations, which range from $.6<C<1.1$ and $4<\alpha<10$ as seen in Figure~\ref{fig:profile_parameter_stats_final_v1}. This aligns with prior works that find $C = 0.75$ \citep{Essig_2019}.

\section{Conclusion}\label{sec:conclusions}

In this work, we have introduced an extension to the parametric model of \cite{yang2024parametric} which takes into account the effects of mass accretion onto SIDM halos. The only change relative to the original parametric model is that we modify how the dimensionless time parameter, $\tau$, is computed, by including a term proportional to the mass accretion rate in the halo. Qualitatively, this model assumes that mass growth drives the evolution of a SIDM halo back toward the CDM limit.

We calibrate and test this model using matched pairs of cosmological zoom-in simulations, one using SIDM physics and another using CDM physics. The CDM N-body simulation is used as the input to the parametric model, which then predicts the properties of the corresponding SIDM halos. The matched SIDM N-body simulation is then used to check the success of the model by examining distributions of properties such as $V_{\max}$ and halo density profiles. We find that the extended parametric model performs substantially better in predicting $V_\mathrm{max}$ for SIDM halos than the original parametric model. In particular, it largely negates a frequent failure mode of the original parametric model in which $V_\mathrm{max}$ is substantially overestimated, which led to halos being predicted to enter core collapse too soon. Predictions for density profiles can be further improved by adjusting the parameter combination $(C, \alpha)$ used in the extended model. Not surprisingly, density profiles are more difficult to predict than a summary property such as $V_\mathrm{max}$. For example, the effects of minor and major mergers are not fully captured even by our extended parametric model due to the forms of Equations~(\ref{eq:param_Vmax_cal} and \ref{eq:param_Rmax_cal}), which are calibrated to perfectly isolated SIDM halos. A more detailed understanding of how density profiles may change due to mergers is left for future work, with which we could also seek to help further extend the parametric model to better predict the gravothermal core collapse of sub-halos along with field halos. 

\section*{Acknowledgments}
We thank Haibo Yu and Daneng Yang for discussions which helped to improve this work. This work uses data from the SIDM Concerto simulations \citep{Nadler_2025_concerto}, which are publicly available at Zenodo at DOI: \href{https://doi.org/10.5281/zenodo.14933624}{10.5281/zenodo.14933624}.
\appendix

\section{Formation Time} \label{sec:formation_time_app}

In both the original and extended parametric model, the formation time, $t_\mathrm{f}$, of halos is crucial since it determines when to start the integrals of Equations~\ref{eq:vmaxintegral} and \ref{eq:rmaxintegral} for the parametric model. $t_\mathrm{f}$ is calculated by first finding the formation redshift using \citep{correa2015accretion}:
\begin{eqnarray}
    z_\mathrm{f} &=& -0.0064 \left[ \log_{10} \left( \frac{M_{\mathrm{vir},0}}{10^{10} \, \mathrm{M}_\odot} \right) \right]^2 \nonumber \\
    & &- 0.1043 \log_{10} \left( \frac{M_{\mathrm{vir},0}}{10^{10} \, \mathrm{M}_\odot} \right) + 1.4807,
\end{eqnarray}
and then simply converting to $t_\mathrm{f}$ using the cosmological time-redshift relation.

\section{Calculating SIDM Profile Parameters in the Parametric Model}\label{sec:SIDM_param_app}

\cite{yang2024parametric} attempt to predict $\rho_\mathrm{s}$, $r_\mathrm{s}$, and $r_\mathrm{c}$ profile parameters for Equation~\ref{eq:sidm_profile} by using $V^\mathrm{(SIDM)}_\mathrm{max}(t)$ and $R^\mathrm{(SIDM)}_\mathrm{max}(t)$ as predicted by the parametric model using Equations~\ref{eq:vmaxintegral} and \ref{eq:rmaxintegral}. This is done by computing $V^\mathrm{(CDM)}_\mathrm{max,0}$ and $R^\mathrm{(CDM)}_\mathrm{max,0}$ for a \emph{CDM halo that would lead to the SIDM halo lying on the same function of $\tau$ as was calibrated to the isolated halos}. That is, they solve Equations~\ref{eq:param_Vmax_cal} and \ref{eq:param_Rmax_cal} for  $V^\mathrm{(CDM)}_\mathrm{max,0}$ and $R^\mathrm{(CDM)}_\mathrm{max,0}$ by substituting $V^\mathrm{(SIDM)}_\mathrm{max}(t)$, $R^\mathrm{(SIDM)}_\mathrm{max}(t)$, and $\tau$.

At every timestep this essentially creates a hypothetical CDM halo at the formation time with properties $V^\mathrm{(CDM)}_\mathrm{max, 0}$ and $R^\mathrm{(CDM)}_\mathrm{max, 0}$,  that would result in the correct $V^\mathrm{(SIDM)}_\mathrm{max}(t)$ and $R^\mathrm{(SIDM)}_\mathrm{max}(t)$ with Equations~\ref{eq:param_Vmax_cal} and \ref{eq:param_Rmax_cal}. Then $V^\mathrm{(CDM)}_\mathrm{max, 0}$ and $R^\mathrm{(CDM)}_\mathrm{max, 0}$ are used to determine $\rho_{s,0}^\mathrm{(CDM)} = \mathrm{G}^{-1} \left(V_{\max,0}^{\text{(CDM)}}/1.648r_\mathrm{s,0}^\mathrm{(CDM)}\right)^2$ and $r_\mathrm{s,0}^\mathrm{(CDM)} = R_{\max,0}^{\text{(CDM)}}/2.1626$ as appropriate for an NFW density profile. In turn, these CDM density profile parameters, along with $\tau$, are used to find the corresponding parameters of the SIDM density profile using the fitting functions given by \cite{yang2024parametric} for isolated SIDM halos:

\begin{eqnarray}
\frac{\rho_\mathrm{s}^{\text{SIDM}}(t)}{\rho_\mathrm{s,0}^\mathrm{(CDM)}} &=&
2.033 + 0.7381\tau + 7.264\tau^5 - 12.73\tau^7 + 9.915\tau^9 \nonumber \\
& & + (1 - 2.033) \frac{\ln(\tau + 0.001)}{\ln 0.001},
\label{eq:profile_model_cal_rhos}
\end{eqnarray}

\begin{eqnarray}
\frac{r_\mathrm{s}^{\text{SIDM}}(t)}{r_\mathrm{s,0}^\mathrm{(CDM)}} &=&
0.7178 + 0.1026\tau + 0.2474\tau^2 - 0.4079\tau^6 \nonumber \\
& & + (1 - 0.7178) \frac{\ln(\tau + 0.001)}{\ln 0.001},
\label{eq:profile_model_cal_rs}
\end{eqnarray}
and
\begin{equation}
\frac{r\mathrm{c}^{\text{SIDM}}(t)}{r_\mathrm{s,0}^\mathrm{(CDM)}} =
2.555\sqrt{\tau} - 3.632\tau + 2.131\tau^2 - 1.415\tau^3 + 0.4683\tau^4.
\label{eq:profile_model_cal_rc}
\end{equation}
Thus, at each timestep, we can use $V_{\max,0}^{\text{(CDM)}}$ and $R_{\max,0}^{\text{(CDM)}}$ computed using the parametric model to obtain $\rho_\mathrm{s}$, $r_\mathrm{s}$, and $r\mathrm{c}$ with which we can generate a density profile for the SIDM halo given by Equation~\ref{eq:sidm_profile}.

\section{Central Density}

We also compare the central density that the parametric models predict with the central densities of the halos from the SIDM N-body simulations at $z=0$. As in Section \ref{sec:results_group} and \ref{sec:results_MW}, we show the results for the extended parametric model versus each of the SIDM N-body simulations, Group and Milky Way, separately. In this paper, we choose to measure the central density of the SIDM halos by calculating the density of the inner 200 particles, which we call $\rho^\mathrm{SIDM}_{c}$. We also record the radius, $r_{200}$, defined as the radius of the 200$^\mathrm{th}$ innermost particle, and use that radius to calculate the central density from the parametric model by using the SIDM profile produced from the parametric model. This is done by taking the parameters $\rho_s$, $r_c$, and $r_c$ that the extended parametric model has calculated for $z=0$. We then obtain the extended model predicted SIDM profile by plugging the three parameters into Equation~\ref{eq:sidm_profile}. Once we have the SIDM profile predicted by the extended parametric model, we then integrate it from $r = 0$ to $r = r_{200}$ to find the total mass withing $r_{200}$ and finally divide by the volume within $r_{200}$ to obtain the central density predicted by the parametric model, $\rho_{c, \mathrm{model}}$. Similar to Equations~\ref{eq:Vmax_percent_error} and \ref{eq:Rmax_percent_error}, we express how well the parametric model predicts the central density with 

\begin{equation}
    \delta_{\rho_c} = \frac{\rho_{c, \mathrm{model}}(z=0) - \rho^\mathrm{(SIDM)}_c(z=0)}{\rho^\mathrm{(SIDM)}_c(z=0)}.
    \label{eq:central_density_percent_error}
\end{equation}

For a given combination of $(C, \alpha)$, we run the extended parametric model for every halo in our Group sample and calculate its $\delta_{\rho_c}$. For that particular run of the extended parametric model using $(C, \alpha)$ we calculate the mean for all $\delta_{\rho_c}$, $\langle \delta_{\rho_c} \rangle$ and the root mean square, $\mathrm{RMS}(\delta_{\rho_c})$. 

Before we discuss the results, it is essential to note that using the inner 200 particles to estimate the central density is an approximation---it is difficult to determine the radius that encompasses the core or center of a SIDM halo, thus, for some halos in the N-body simulation, the inner 200 particles may be insufficient to accurately capture the central density. 

\begin{figure*}[b]
\begin{center}
\begin{tabular}{cc}
    \includegraphics[width=0.45\linewidth]{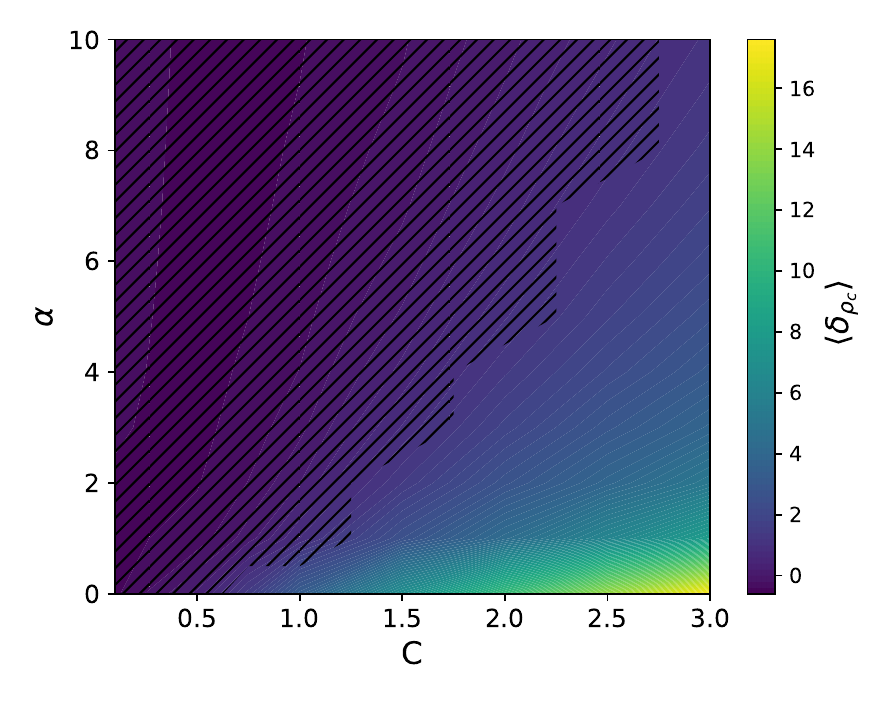}  &  \includegraphics[width=0.45\linewidth]{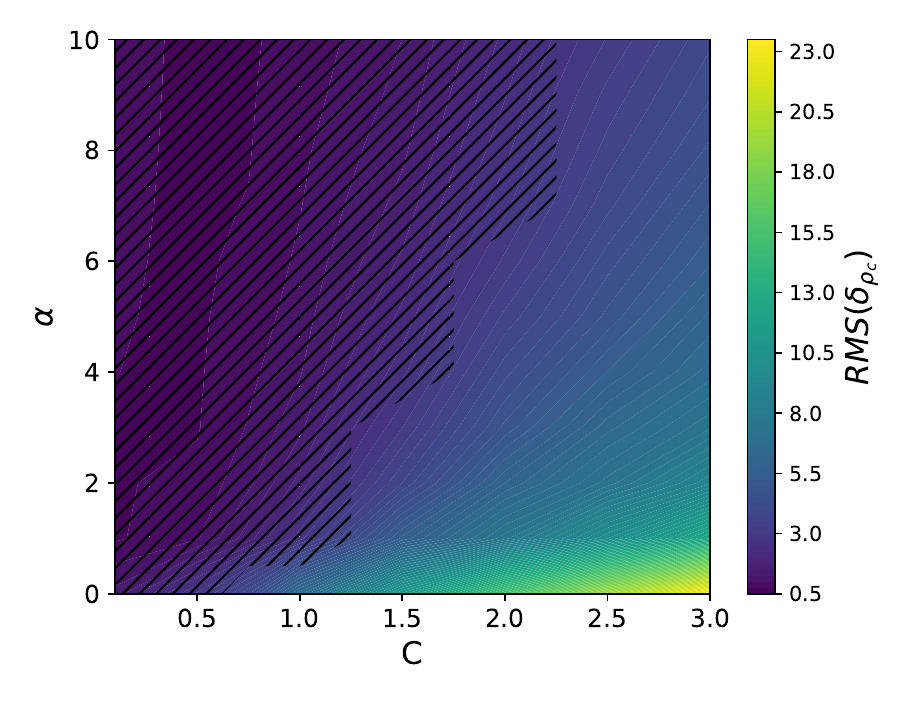}
\end{tabular}
\end{center}
\caption{2D color maps measuring the performance of the extended parametric model in predicting central densities of SIDM halos in the Group simulation. The left panel shows $\langle \delta_{\rho_c} \rangle$ for a range of $C$ and $\alpha$ for the extended parametric model, while the right panel shows $\mathrm{RMS} (\delta_{\rho_c})$. The hatched area depicts the region of $(C, \alpha)$ parameters for which the extended parametric model performs better than the original parametric model.}
\label{fig:cd200_group}
\end{figure*}

\begin{figure*}
\begin{center}
\begin{tabular}{cc}
    \includegraphics[width=0.45\linewidth]{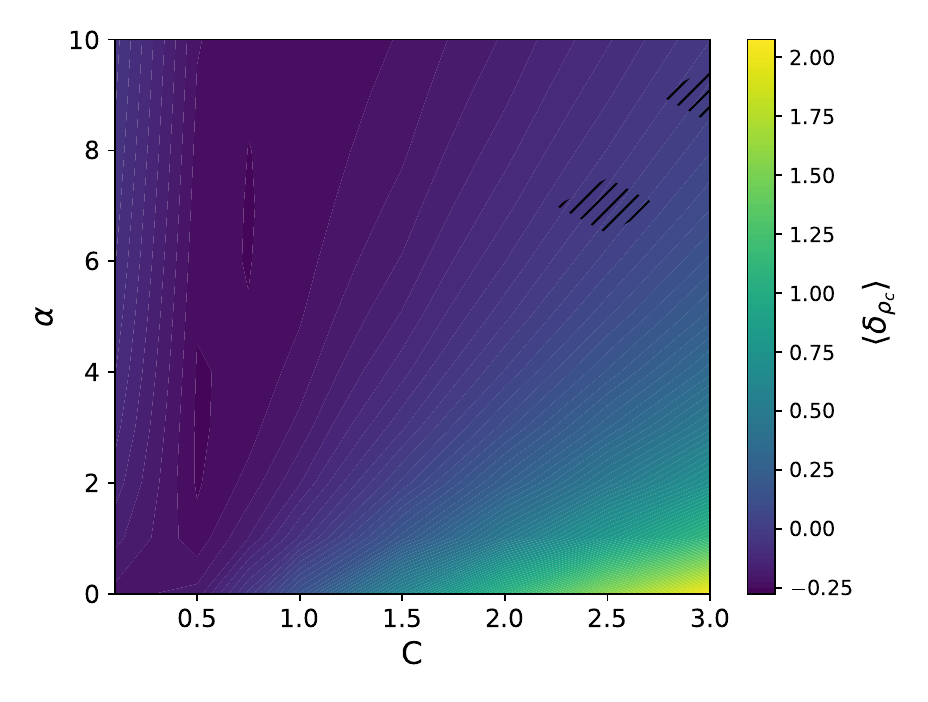}  &  \includegraphics[width=0.45\linewidth]{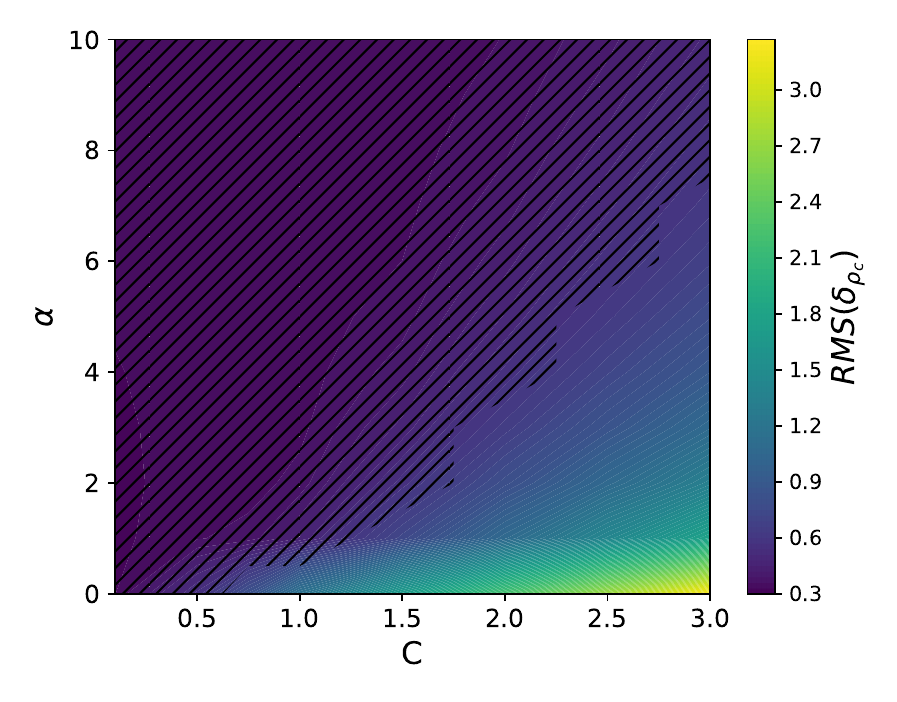}
\end{tabular}
\end{center}
\caption{2D color maps measuring the performance of the extended parametric model in predicting central densities of SIDM halos in the Milky Way simulation. The left panel shows $\langle \delta_{\rho_c} \rangle$ for a range of $C$ and $\alpha$ for the extended parametric model, while the right panel shows $\mathrm{RMS} (\delta_{\rho_c})$. The hatched area depicts the region of $(C, \alpha)$ parameters for which the extended parametric model performs better than the original parametric model.}
\label{fig:cd200_final}
\end{figure*}

We create 2D color maps to show which values of $(C, \alpha)$ the extended model best predicts the central density in Figure~\ref{fig:cd200_group}. The hatched region shows which $(C, \alpha)$ parameter values predict $\delta_{\rho_c}$ better than the original model. For the Group simulation, the 2D color maps in Figure~\ref{fig:cd200_group} are similar to Figure~\ref{fig:profile_parameter_stats_group_v1}. We can see that the extended model does predict central density better than the original model, especially for values $C<1.5$, because $\langle \delta_{\rho_c} \rangle$ and $\mathrm{RMS}(\delta_{\rho_c})$ have the lowest values when $C<1.5$. For the Milky Way simulation, we once again see tighter constraints on what values of $(C, \alpha)$ are allowed compared to the Group simulation. Figure~\ref{fig:cd200_final} shows $\langle \delta_{\rho_c} \rangle$ and $\mathrm{RMS}(\delta_{\rho_c})$ for the Milky Way population of halos. On the right panel we see the $\mathrm{RMS}(\delta{\rho_c})$ follow similar patterns to Figures~\ref{fig:profile_parameter_stats_group_v1}, \ref{fig:profile_parameter_stats_final_v1}, and \ref{fig:cd200_group} where the extended model tightens the spread of the $\delta$ statistic we are looking at over a region of $(C, \alpha)$ that suggests that to lower $\mathrm{RMS}$ values one must decrease $\alpha$ as $C$ increases. The left panel is where we see that, on average across the Milky Way sample of field halos, the extended model struggles to perform better than the original in predicting central densities except for a handful of combinations of $(C,\alpha)$ values. It is important to note that, depending on the number of inner particles we choose to represent the central density, Figure~\ref{fig:cd200_final} changes, and the values of $(C,\alpha)$ for which the extended model predicts the central density better than the original model change. We recommend interpreting Figures~\ref{fig:cd200_group} and \ref{fig:cd200_final} qualitatively, with the main takeaway being that the extended model performs better at predicting central density for the Group sample of field halos. 

\clearpage

\bibliography{biblio}
\bibliographystyle{aasjournal}

\end{document}